\documentclass[a4paper,onecolumn,11pt,accepted=2023-08-10]{quantumarticle}
\pdfoutput=1
\usepackage{geometry}                
\usepackage{graphicx}
\usepackage{subcaption}
\usepackage{fullpage}
\usepackage{amssymb}
\usepackage{epstopdf}
\usepackage{amsthm}
\usepackage{amsmath}
\usepackage{thmtools}
\usepackage{thm-restate}
\usepackage{bm}
\usepackage{hyperref}
\usepackage[natbib=true,style=alphabetic,backend=bibtex,isbn=false,url=false,maxbibnames=6,maxcitenames=2,useprefix=true]{biblatex}
\usepackage[normalem]{ulem}
\makeatletter
\def\th@plain{%
  \thm@notefont{}
  \itshape 
}
\def\th@definition{%
  \thm@notefont{}
  \normalfont 
}
\makeatother

\usepackage{amsfonts}
\usepackage{mathtools}
\usepackage{bigints}
\usepackage{xcolor}
\usepackage{enumitem}
\usepackage{afterpage}
\usepackage{setspace}
\usepackage{verbatim}
\usepackage{physics}

\usepackage{braket}

\newtheorem{theorem}{Theorem}[section]
\newtheorem{proposition}{Proposition}[section]

\newtheorem{lemma}{Lemma}[section]

\theoremstyle{definition}

\newcommand{\Var}{\mathrm{Var}}
\newcommand{\Cov}{\text{Cov}}
\newcommand{\Poi}{\text{Poi}}

\newcommand{\SWdens}[2]{\mathrm{SW}^{#1}_{#2}}
\newcommand{\SWunif}[2]{\mathrm{SW}^{#1}_{#2}}

\begin{document}
\title{Testing identity of collections of quantum states: sample complexity analysis}

\author{Marco Fanizza}
\affiliation{F\'{\i}sica Te\`{o}rica: Informaci\'{o} i Fen\`{o}mens Qu\`{a}ntics, Departament de F\'{\i}sica, Universitat Aut\`{o}noma de Barcelona, 08193 Bellaterra, Spain.}
\email{marco.fanizza@uab.cat} 
\orcid{0000-0003-0802-8000}
\author{Raffaele Salvia}
\affiliation{Scuola Normale Superiore, I-56127 Pisa, Italy.} 
\orcid{0000-0002-0006-7630}
\email{raffaele.salvia@sns.it}
\author{Vittorio Giovannetti} \affiliation{NEST, Scuola Normale Superiore and Istituto Nanoscienze-CNR, I-56127 Pisa, Italy.}
\orcid{0000-0002-7636-9002}
\maketitle
\begin{abstract}
We study the problem of testing identity of a collection of unknown quantum states given sample access to this collection, each state appearing with some known probability. We show that for a collection of $d$-dimensional quantum states of cardinality $N$, the sample complexity is $O(\sqrt{N}d/\epsilon^2)$, {with a matching lower bound, up to a multiplicative constant}. The test is obtained by estimating the mean squared Hilbert-Schmidt distance between the states, thanks to a suitable generalization of the estimator of the Hilbert-Schmidt distance between two unknown states by Bădescu, O'Donnell, and Wright (https://dl.acm.org/doi/10.1145/3313276.3316344).
\end{abstract}
\tableofcontents

\section{Introduction}

The closeness between quantum states can be quantified according to a variety of unitarily invariant distance measures, with different operational interpretations~\cite{Hayashi2016}. Given access to copies of some unknown states, a fundamental inference problem is to tell if the states are equal or distant more than $\epsilon$, according to some unitarily invariant distance. Since this problem does not require to completely reconstruct the unknown states with tomography protocols, optimal algorithms require less copies than full tomography to answer successfully~\cite{Badescu2019}. Due to unitarily invariance, efficient algorithms can also be guessed by symmetry arguments~\cite{Hayashi2017}.

In this work we study the problem of testing identity of a collection of unknown quantum states given sample access to the collection. We show that for a collection of $d$-dimensional quantum states of cardinality $N$, the \textit{sample complexity} is $O(\sqrt{N}d/\epsilon^2)$, which is optimal up to a constant. We assume a \textit{sampling model access}, where each state appears with some known probability, adapting~\cite{levi2013testing,Diakonikolas2016} to the quantum case. {We also consider a Poissonized version of the sampling model, where the number of each copies of a state is a Poissonian random variable, and we show that the sample complexity of the two models is the same.}  
This problem is an example of \textit{property testing}, a concept developed in computer science~\cite{Goldreich2017}, and applied to hypothesis testing of distributions~\cite{Canonne2020} and quantum states and channels~\cite{Montanaro2016}. At variance with optimal asymptotic error rates studied in statistical classical and quantum hypothesis testing~\cite{lehmann2006testing, Hayashi2016}, the sample complexity captures finite size effect in inference problems, as it expresses the number of samples required to successfully execute an inference task in terms of the extensive parameters of the problem, in our case the dimension $d$ and the cardinality $N$ of the collection. The interest in these kind of questions in the classical case has been motivated by the importance of the study of big data sources; a similar motivation holds for the quantum case, since outputs of fully functional quantum computers will also live in high-dimensional spaces.

\subsection{Results}\label{sec:res} 

Given a collection of $d$-dimensional quantum states $\{\rho_i\}_{i=1,...,N}$, and a probability distribution $p_i$ ($0<p_i<1$), we consider a \textit{sampling model}~\cite{levi2013testing,Diakonikolas2016} where we have access to {{$M$}} copies of the density matrix
\begin{eqnarray}
\label{stato_sampling}
\rho= \sum_{i=1}^N p_i \ketbra{i}{i} \otimes \rho_i,
\end{eqnarray}
where $\{\ket{i}\}_{i=1,...,N}$ is an orthonormal basis of a $N$ dimensional (classical) register. We are promised that one of the two following properties holds:

\begin{itemize}
\item {\bf Case $\mathsf{A}$}: $\rho_1=\rho_2=...=\rho_N$,  which can be equivalently stated by saying that there exists a $d$-dimensional state   $\sigma$ such that 
		$\sum_{i} p_i D_{\mathrm{Tr}}(\rho_i,\sigma)=0$, with $D_{\mathrm{Tr}}$ the {{trace distance~\cite{Hayashi2016}}}; 
\item {\bf Case $\mathsf{B}$}: For any $d$-dimensional state $\sigma$ it holds $\sum_{i} p_i D_{\mathrm{Tr}}(\rho_i,\sigma)>\epsilon$.
\end{itemize}

	Our goal is to find the {{values of $M$ for which there is a two-outcome test that}}  can discriminate the two cases 
	with  high probability  of success. Explicitly, indicating with  \texttt{"accept"}  and  \texttt{"reject"} the outcomes of the test, we 
	require
	the probability of getting \texttt{"accept"} to be larger than  $2/3$ in case $\mathsf{A}$, and smaller than 
	$1/3$ in case $\mathsf{B}$, i.e. 
	\begin{eqnarray} 
	\left\{ \begin{array}{l}  \label{constraint} 
	P( \mbox{\texttt{test $\mapsto$ "accept"} }| \mbox{Case $\mathsf{A}$}  ) >  2/3 \;, \\ \\
	P( \mbox{\texttt{test $\mapsto$ "accept"} }|  \mbox{Case $\mathsf{B}$} )<  1/3\;. \end{array} \right.
	\end{eqnarray}  Note that the values $2/3$ and $1/3$ are by convention and, { as long as we are interested in the sample complexity only up to a scaling factor,} can be replaced by any pair of constants $c,s$, respectively, { such that $1>c>s>0$}. The main result of the paper is to provide an estimate of necessary and sufficient values of $M$ to fulfill the above conditions. 
We use the notations $O(f(d,N,\epsilon))$ and $\Omega(g(d,N,\epsilon))$ to indicate respectively upper and lower bounds to sample complexities, up to multiplicative constants. If lower and upper bounds which differ by a multiplicative constant can be obtained, the sample complexity is considered to be determined and indicated as $\Theta(f(d,N,\epsilon))=\Theta(g(d,N,\epsilon))$.

Specifically we prove the following results:

\begin{theorem}\label{theoremtr}
{For any $\epsilon>0$}, given access to $O\left(\frac{\sqrt{N}d}{\epsilon^2}\right)$ samples of the density matrix $\rho$ of Eq.~(1), there is an algorithm which can distinguish with high probability whether 
\begin{itemize}
\item $\sum_{i} p_i D_{\mathrm{Tr}}(\rho_i,\sigma)> \epsilon$ { for every state $\sigma$ {(Case $\mathsf{B}$)}, or
\item there exists a state $\sigma$ such that} $\sum_{i} p_i D_{\mathrm{Tr}}(\rho_i,\sigma)=0$ { (that is, all the states  $\rho_i$ are equal, {Case $\mathsf{A}$})}.
\end{itemize}
\end{theorem}

\begin{theorem}\label{theoremlow}
{For any $\epsilon>0$}, any algorithm which can distinguish with high probability whether 
\begin{itemize}
\item $\sum_{i} p_i D_{\mathrm{Tr}}(\rho_i,\sigma)> \epsilon$ { for every state $\sigma$ {(Case $\mathsf{B}$)}, or 
\item there exists a state $\sigma$ such that} $\sum_{i} p_i D_{\mathrm{Tr}}(\rho_i,\sigma)=0$ { (that is, all the states  $\rho_i$ are equal, {Case $\mathsf{A}$})},
\end{itemize} 
given access to $M$ copies of the density matrix $\rho$ of Eq.~(1), requires at least $M=\Omega\left(\frac{\sqrt{N}d}{\epsilon^2}\right)$ copies.
\end{theorem}

The proof of Theorem~\ref{theoremlow} is presented in Sec.~\ref{Sec:lower} and it relies on  the fact that a test working with $M$ copies could be used to discriminate between two states which are close in trace distance unless $M=\Omega\left(\frac{\sqrt{N}d}{\epsilon^2}\right)$. These states are obtained as average inputs $\rho_{A}$ and $\rho_{B}$ of the form of Eq.~(\ref{stato_sampling}) for two different set of collections of states: in the first case the set is made of only one collection consisting of maximally mixed states (thus satisfying case $\mathsf A$), and in the second the set of collections is such that its elements satisfy case $\mathsf B$ with high probability. {The technical contributions of this proof are (a) a lower bound on the probability that a collection of random states with spectrum $s_{\epsilon}=(\frac{1+\epsilon}{d},\frac{1-\epsilon}{d},...,\frac{1+\epsilon}{d},\frac{1-\epsilon}{d})$ has large average trace distance to their average state; (b) an upper bound on the distance between $\rho_A$ and $\rho_B$ being the average input state over collections of random states with spectrum $s_{\epsilon}$. Both results could be useful elsewhere.}

The derivation of the upper bound for $M$  given in Theorem~\ref{theoremtr}  is  
instead presented in Sec.~\ref{upbound} and it is 
obtained by constructing an observable $\mathcal{D}_M$ whose expected value is the mean squared Hilbert-Schmidt distance between the states $\rho_i$, and we bound the variance of the estimator. By relating the mean squared Hilbert-Schmidt distance to $\sum_{i} p_i D_{\mathrm{Tr}}(\rho_i,\sum_{i}p_i\rho_i)$ we obtain the test of the theorem. {This strategy follow closely the methods of~\cite{Badescu2019} (for $N=2$), although with some relevant changes due to the fact that we are not requiring a fixed number of copies of each state {$\rho_i$}, like in~\cite{Badescu2019}. This difference is relevant from a conceptual point of view, since having an arbitrary number of copies of each state is a stronger type of access with respect to the sampling model, and closer to the query model (we discussed the different applicability scenario in the following section). It is also relevant from a technical point of view, since it is not immediate to devise an estimator for which the analysis can be completed.}
In fact, the analysis exploits a {\it Poissonization} trick~\cite{levi2013testing} where the number of copies $M$ is not fixed but a random variable, extracted from a Poisson distribution with average $\mu$, $\Poi_{\mu}(M):=\frac{e^{-\mu}\mu^M}{M!}$ {(summarized later on by the notation $M\sim \Poi_\mu$)}. We then look for a test which can be performed by a two-outcome POVM $\{E_0^{(M)},E_1^{(M)}\}$ for each $M$.  Poissonization is a standard technique that allows the for some useful simplification of the analysis by  getting rid of unwanted  correlations (more on this in Sec.~\ref{subsec31}).
The equivalence of the Poissonized model with the original one is formalised in Appendix~\ref{Poissonapp}.

{
Analogously to~\cite{Badescu2019} we can refine the upper bound when the states in the collection have low rank. 
Given the state $\rho$ of Eq.~(\ref{stato_sampling}), we define its reduced average density matrix  
\begin{eqnarray} \bar \rho:=\sum_{i=1}^Np_i\rho_i\;, \label{def:avstate}\end{eqnarray}

In particular, when $\bar\rho$ is $\eta$-close to rank $k$, that is, the sum of its $k$ largest eigenvalues is larger than $1-\eta$, we can refine Theorem~\ref{theoremtr}:
\begin{theorem}\label{theoremtr_rangok}
If the density matrix $\bar\rho$ of Eq.~(\ref{def:avstate}) is $\eta$-close to rank $k$, given access to $O\left(\frac{\sqrt{N}k}{\epsilon^2}\right)$ samples of $\rho$ there exists an algorithm which can distinguish with high probability whether $\sum_{i} p_i D_{\mathrm{Tr}}(\rho_i,\sigma)> \epsilon + \eta$ { for every state $\sigma$, or there exists a state $\sigma$ such that} $\sum_{i} p_i D_{HS}(\rho_i,\sigma) <8 (2 - \sqrt{2})\epsilon$.
\end{theorem}

}

{ 
\subsection{Motivation of the setting}
In this section we present a couple of physical settings which give rise to the sampling models discussed in Sec.\ref{sec:res}, as both the original model and the Poissonized model refer to natural scenarios for a certification task. 

\textbf{ Independent sources setting} (panel (a) of Figure~\ref{settings}). It is fair to assume that each copy of the states is produced by a device $\mathcal{S}_i$ that require some physical time to run, and produces the expected state with some probability. Moreover, assume that the number of produced copies of $\rho_i$ by $\mathcal{S}_i$ at any time $T$ is given by a Poisson distribution with rate $r_i$ and average $r_iT$, i.e. $P_T(m_i)=\Poi_{r_i T}(m_i)=\frac{(r_i T)^{m_i}e^{r_i T}}{m_i!}$. With this assumption, it also holds that the probability that a total of $M=m_1+...+m_N$ copies is produced in the time $T$ is  $P_T(M)=\frac{(T\sum_{i=1}^Nr_i)^{M}e^{T\sum_{i=1}^N r_i}}{M!}$. The Poissonized sampling model (where the probabilities of getting $m_i$ copies of $\rho_i$ are given by a Poisson distribution with average $p_i\mu$, see Eq.~(\ref{POISIDE})) is an adequate representation of the setting where we want to do our certification test with all the copies that are produced in a certain timeframe $T$, see panel (a) of Figure~\ref{settings}.

On the other hand, if we decide to run the test as soon the total number of copies corresponds to the desired number $M$, we end up in the original sample model. Indeed, if $T_{M}$ is the random variable equal to the time at which the total number of copies is $M$, we have that the probability of finding a vector $\vec m=(m_1,...,m_N)$ of number of copies $\rho_{1},...,\rho_{N}$, respectively, conditioned on $m_1+...+m_N=M$ at the time $T_{M}$, is
\begin{equation}
P(\vec{m}|M)=\int_{0}^{\infty}p(T_M=T) P_T(\vec{m}|M)dT,\\
\end{equation}
where $p(T_M=T)$ is the probability density for the stopping time $T_M$, and
\begin{align}
 P_T(\vec{m}|M)&= P_T(\vec{m},M)/P_T(M)=  P_T(\vec{m})/P_T(M)= \prod_{j=1}^{N}P_T(m_j)/P_T(M)\\
 &=\frac{M!}{m_1!...m_N!} \prod_{j=1}^{N} r_i^{m_i}.
\end{align}
where the first equality comes from the definition of conditional probability, the second comes from the fact that $M$ is completely determined by $\vec{m}$, the third comes from the fact that the components of $\vec{m}$ are independent when conditioning only on $T$, and the last equality comes from writing the probabilities explicitly. Finally, by integrating a constant function, we have
\begin{equation}
P(\vec{m}|M)=\frac{M!}{m_1!...m_N!} \prod_{j=1}^{N} r_i^{m_i},
\end{equation}
which is the probability distribution of the copies of each $\rho_i$ in the original sampling model with $M$ total copies of $\rho$, provided that $r_i=p_i$.

These two situations can be compared with the setting of the \textit{query model}, already considered in \cite{yu2020sample}; in that case, we are allowed to ask for any number of copies of each state {$\rho_i$} in the collection, and the sample complexity is measured with respect to the total number of copies requested. This type of access is clearly stronger with respect to the sampling models, and indeed the sample complexity is lower, being $\Theta(d/\epsilon^2)$. However, assuming there is a finite rate of copies/time, the sampling model captures better the actual physical time required to generate the copies for the test. 

On the other hand, the validity of the assumption that the number of copies of each state is generated by a Poisson distribution can be questioned. By the law of rare events, this is a realistic approximation if each source actually corresponds to many independent sources, each of which produces a copy of the state with very small probability, such that the total rate of production of state is finite. In particular, the following bound on the variational distance between the Poisson distribution and sum of independent Bernoulli random variables $X_i\sim (p_i,1-p_i)$ holds~\cite{le1960approximation}: $\sum_{k=0}^{\infty }|P(\sum_{i=1}^n X_i=k)-\frac{(\sum_{i=1}^{n} p_i)^ke^{-(\sum_{i=1}^{n} p_i)}}{k!}|<2(\sum_{i=1}^{n}p_i^2)$. The approximation with i.i.d. Bernoulli variables was considered, for example, for entanglement certification of single-photon pairs produced with spontaneus parametric down conversion~\cite{Hayashi_2008} in the asymptotic setting, with proposed tests implemented experimentally~\cite{PhysRevA.74.062321}. In these cases the single-photon pairs are produced with a very small probability from a single beam, but with a finite rate if the number of beams is large, and the distribution of the total number of pairs is approximated by a Poisson distribution. In any case, since any probabilistic model can be simulated or can simulate our sampling model, simply simulating the desired probability distribution on a classical computer and waiting for enough copies, our protocol gives respectively upper or lower bounds on the sample complexity. These bounds are tight if the simulation is efficient, that is it requires the same number of copies, up to a constant multiplicative factor. It would be interesting to characterize which sampling models can efficiently simulate or be simulated by the Poissonized model, but we will not discuss this issue here.

{\textbf{Noisy measurement setting}} (panel (b) of Figure~\ref{settings}).
We point out another setting where the sampling model can represent a realistic situation in the lab: suppose that some preparation procedure $\mathcal{S}$ ends with some measurement, but different outcomes of the measurement are expected to correspond to the same desired state. An example could be the case if our preparation apparatus has interacted with an environment, and we measure the environment. Since the outcome of the measurement at the preparation stage is random, the procedure prepares in principle different states for each measurement outcome. The classical-quantum state we obtain, possibly after post-selection of acceptable measurement outcomes, will have the form in Eq.~(\ref{stato_sampling}). The goal of the test is to certify if a source of states of this kind is stable or not. 

\begin{figure}
\center
\begin{subfigure}{0.45\textwidth}\includegraphics[width=\textwidth]{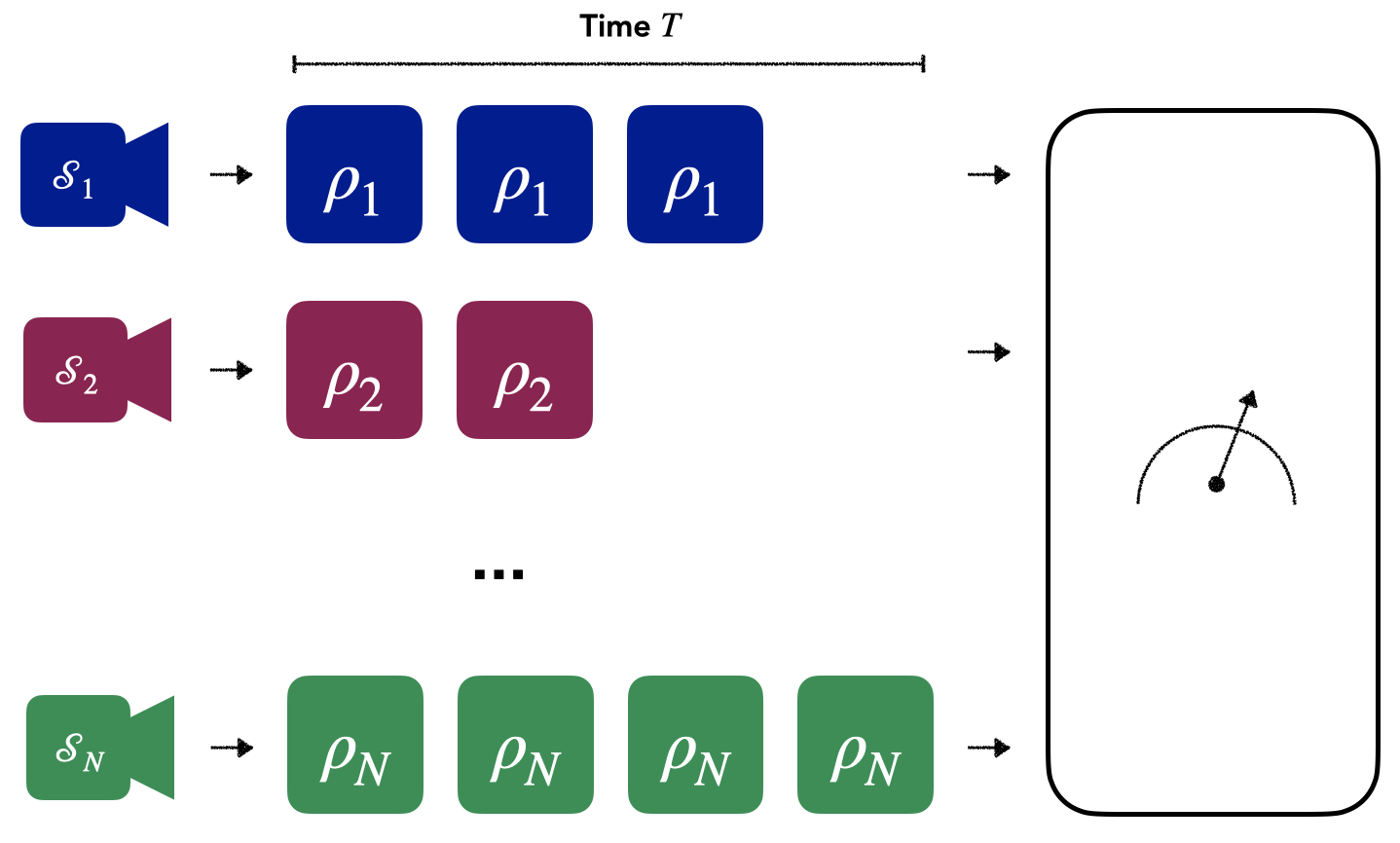}
\caption{}
\end{subfigure}
  \hspace*{0.1\textwidth}   
\begin{subfigure}{0.40\textwidth}\includegraphics[width=\textwidth]{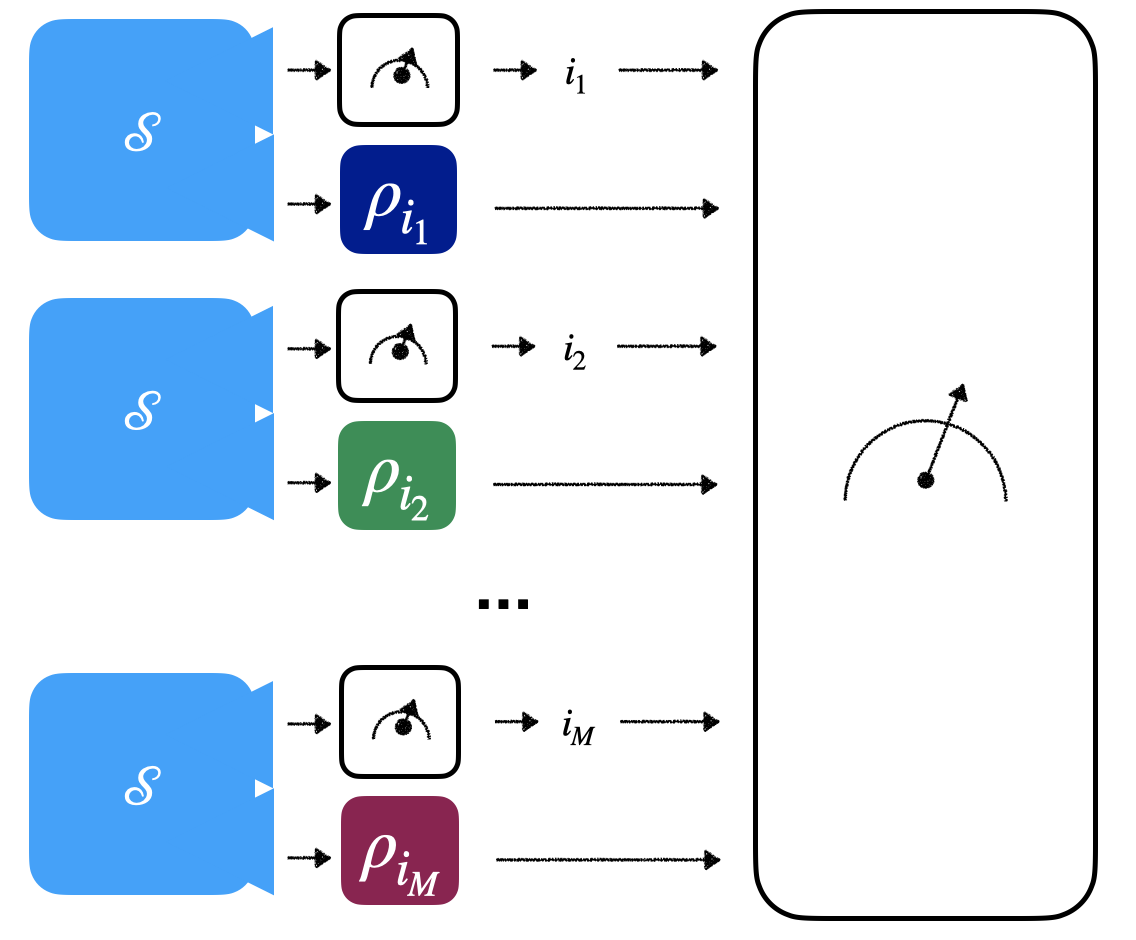}
\caption{}
\end{subfigure}
\caption{\label{settings}Two settings to which our model applies. In a) independent sources produce a random number, Poisson distributed, of copies of a state in a time $T$. In b), a measurement procedure prepares labeled states, each label appearing with some probability.}
\end{figure}

Finally, we point out similarities with the problem of quantum change point detection~\cite{Akimoto2011,Sentis2016,Sentis2017,Sentis2018, Fanizza2022a}, in which a sequence of unknown states is presented, and it is asked if they are all equal or not. An additional question is to identify the points where the states change. With our algorithm, we are able to answer correctly to the change point problem when we know that there could be a change point among $N-1$ possible change points, which have distances between them given by Poissonian random variables. It would be interesting if the analysis and the techniques of the present paper could be extended to address  the change point problem more directly.

}

\subsection{Related work}

\subsubsection{Classical distribution testing}
For an overview of learning properties of a classical distribution in the spirit of property testing, we refer to~\cite{Goldreich2017, Canonne2020}. We report a partial list of results which are of direct interest for this paper, about testing symmetric properties of distribution in total {variation} distance. We use the notation $[d]$ for the set $\{1,...,d\}$. Learning a classical distribution over $[d]$ in total variation distance can be done in $O(d/\epsilon^2)$ samples~\cite{Goldreich2017}, therefore the interest in testing properties is to get a sample complexity $o(d)$. The problem of testing uniformity was addressed in~\cite{Goldreich2011} and established to be $O(\sqrt{d}/\epsilon^2)$ in successive works~\cite{Paninski2008, Valiant2014}. More generally, the sample complexity of identity testing to a known distribution has been established to be $\Theta(\sqrt{d}/\epsilon^2)$~\cite{Valiant2014, Diakonikolas2015}. Identity testing for two unknown distribution is $\Theta(\max(d^{1/2}/\epsilon^2,d^{2/3}/\epsilon^{4/3}))$~\cite{Chan2014}. The problem of testing identity of collection of $N$ distributions was introduced in the classical case in~\cite{levi2013testing} and solved in \cite{Diakonikolas2016}, obtaining $\Theta(\max(\sqrt{dN}/\epsilon^2,d^{2/3}N^{1/3}/\epsilon^{4/3}))$ for the sampling model, where at each sample the tester receives one of $N$ distributions with probabilty $p_i$,  and $\Theta(\max(\sqrt{d}/\epsilon^2,d^{2/3}/\epsilon^{4/3}))$ for the query model, where the tester can choose the distribution to call at each sample. A problem related to testing identity of collections is testing independence of a distribution on $\times_{i=1}^{l}[n_i]$, which was addressed by~\cite{Batu2001, levi2013testing, Acharya2015} and solved in~\cite{Diakonikolas2016}, which showed a tight sample complexity $\Theta(\max_j(\prod_{i=1}^{l}n_i^{1/2}/\epsilon^{1/2},n_j^{1/3}\prod_{i=1}^{l}n_i^{1/3}/\epsilon^{4/3}))$.

\subsubsection{Quantum state testing}
It has been shown that the reconstruction of the classical description of an unknown state, \textit{quantum tomography}, requires $\Theta(d^2/\epsilon^2)$ copies of the state~\cite{Haah2017, ODonnell2016, ODonnell2017}. These algorithms {often include, as a subroutine, spectrum learning}~\cite{Alicki1988,Keyl2001,Hayashi2002b,Christandl2006, Keyl2006}, which has sample complexity $O(d^2/\epsilon^2)$~\cite{ODonnell2016}, although a matching lower bound is available only for the empirical Young diagram estimator~\cite{ODonnell2015a}. These results have been refined in the case the state is known to be close to a state of rank less than $k$. Quantum entropy estimation has been studied in~\cite{acharya2020estimating}. The property testing approach to quantum properties has been reviewed in~\cite{Montanaro2016}, where it is also shown that testing identity to a pure state requires $O(1/\epsilon^2)$. Testing identity to the maximally mixed state takes $\Theta(d/\epsilon^2)$~\cite{ODonnell2015a}, and the same is true for a generic state and for testing identity between unknown states (with refinements if the state can be approximated by a rank $k$ state)~\cite{Badescu2019}. In~\cite{Badescu2019}, identity testing between unknown states is done by first estimating their Hilbert-Schmidt distance with a minimum variance unbiased estimator, developing a general framework for efficient estimators of sums of traces of polynomials of states. This improves on a simple way to estimate the overlap $\Tr[\rho\sigma]$ between two unknown states, the swap test~\cite{Buhrman2001}, while optimal estimation of the overlap between pure states with average error figures of merit has been addressed by a series of works~\cite{Bartlett2004, Bagan2006, Lindner2006, Gisin2006, Fanizza2020a}. In all of these cases, the algorithms considered are classical post-processing of the measurement used to learn the spectrum of a state, possibly repeated on nested sets of inputs. This measurement can be efficiently implemented, with gate complexity $O(n, \log d, \log 1/\delta)$~\cite{Bacon2006, Harrow2005, Krovi2019}, where $n$ is the number of copies of the state, and $\delta$ is the precision of the implementation. This measurement is relevant for several quantum information tasks, for example in communication (see e.g.~\cite{Hayashi2017, Bennett2014}).
Testing identity of collections of quantum states in the query model has been established to be $\Theta(d/\epsilon^2)$~\cite{yu2019quantum}, while the sampling model complexity was left open and is addressed in this paper. Independence testing is also addressed in~\cite{yu2019quantum}, obtaining a sample complexity $O(d_1 d_2/\epsilon^2)$, which is tight up to logarithmic factors, using the identity test of \cite{Badescu2019} for testing independence of a state on $\mathbb C^{d_1}\otimes \mathbb C^{d_2}$; similar results hold for the multipartite case (see also~\cite{Hayashi2014} for the asymptotic setting {and~\cite{Bai2021} for an application of independence testing to identification of causal structure)}. Besides these optimality results, which are valid if one allows any measurement permitted by quantum mechanics, several results have been obtained in the case in which there are restrictions on the measurements:~\cite{bubeck2020entanglement} shows that the sample complexity for testing identity to the maximally mixed state with independent but possibly adaptive measurements is $\Omega(d^{4/3}/\epsilon^2)$ and $\Theta(d^{3/2}/\epsilon^2)$ for non-adaptive measurements, while the instance optimal case for the same problem is studied in~\cite{chen2021toward};~\cite{Haah2017} shows that the sample complexity for tomography for non-adaptive measurements is $\Omega(d^3/\epsilon^2)$. Algorithms with Pauli measurements only have been considered~\cite{yu2019quantum, yu2020sample}, while a general review of the various approaches with attention to feasibility of the measurement can be found in~\cite{Kliesch2020}. 

\section{Preliminaries}
\subsection{Distance measures for collection of distributions}\label{sec:dist} 

Quantum states are positive operators in a Hilbert space, with trace one. In this work we consider states living in a Hilbert space of finite dimension $d$  {{and we  make use of the Schatten operator norms~\cite{Hayashi2016}: $||A||_p=\Tr[\sqrt{A^\dagger A}^p]^{1/p}$.
In particular, given $\rho$ and $\sigma$ two quantum states of the system,  we express their trace distance  as $D_{\Tr}(\rho,\sigma)$ and their Hilbert-Schmidt distance $D_{\mathrm{HS}}(\rho,\sigma)$  as 
\begin{eqnarray}
D_{\Tr}(\rho,\sigma)=\frac{||\rho-\sigma||_1}{2} \;, \qquad \qquad 
D_{\mathrm{HS}}(\rho,\sigma)=||\rho-\sigma||_2\;. 
\end{eqnarray}
These quantities are connected via the following inequalities
\begin{equation}
\frac{1}{2}D_{\mathrm{HS}}(\rho,\sigma)\leq D_{\Tr}(\rho,\sigma) \leq \frac{\sqrt{d}}{2}D_{\mathrm{HS}}(\rho,\sigma).
\end{equation}
We also recall that the trace distance admits a clear operational interpretation due to the
Holevo-Helstrom theorem~(see e.g. \cite{Hayashi2016}): if a state is initialized as $\rho$ with probability $1/2$ and $\sigma$ with probability $1/2$, the maximum probability of success in identifying the state correctly is given by:}}
\begin{equation}\label{HH}
p_{succ}(\rho,\sigma)=\frac{1}{2}\left(1+D_{\mathrm {Tr}}(\rho,\sigma)\right).
\end{equation}

{{For $\rho$ and $\bar\rho$ as defined in Eq.~(\ref{stato_sampling}) and~(\ref{def:avstate}), we introduce the quantity
\begin{align}
\mathcal{M}_{\Tr}(\rho) &:=\sum_{i=1}^N p_i D_{\Tr}(\rho_i, \bar\rho)\leq \frac{1}{2}\sum_{i=1}^N p_i \sqrt{dD^2_{HS}(\rho_i, \bar\rho)}\;.
\end{align}
We also define the mean squared Hilbert-Schmidt distance of the model as 
\begin{eqnarray}
\mathcal{M}_{HS}(\rho) := \left[\sum_{i=1}^N\sum_{j=1}^N p_i p_j D^2_{HS}\left(\rho_i, \rho_j\right) \right]^{1/2},
\end{eqnarray}
observing that it can be equivalently expressed in terms of $\bar \rho$ as }}
\begin{align}
\mathcal{M}_{HS}^2(\rho) &:=\sum_{i=1}^N\sum_{j=1}^N p_i p_j D^2_{HS}\left(\rho_i, \rho_j\right)=\sum_{i=1}^N\sum_{j=1}^N p_i p_j\Tr[(\rho_i-\rho_j)^2] \nonumber\\
&= \sum_{i=1}^N\sum_{j=1}^N p_i p_j \Tr[(\rho_i-\bar \rho+\bar\rho-\rho_j)^2]\nonumber\\
&= 2\sum_{i=1}^N p_i\Tr[(\rho_i-\bar \rho)^2]-2\sum_{i=1}^N\sum_{j=1}^N p_i p_j\Tr[(\rho_i-\bar \rho)(\rho_j-\bar \rho)]\nonumber \\
&= 2\sum_{i=1}^N p_i D^2_{HS}(\rho_i,\bar \rho).
\end{align}
{{Therefore we can derive the following important inequality 
\begin{align}\label{basic_ineq}
\mathcal{M}_{\Tr}(\rho) &=\sum_{i=1}^N p_i D_{\Tr}(\rho_i, \bar\rho)\leq \frac{1}{2}\sum_{i=1}^N p_i \sqrt{dD^2_{HS}(\rho_i, \bar\rho)} \nonumber \\ 
&\leq \frac{1}{2}\sqrt{ \sum_{i=1}^N p_i} \sqrt{\sum_{i=1}^N p_i dD^2_{HS}(\rho_i, \bar\rho)}
 = \frac{\sqrt{d}}{2\sqrt{2}}\mathcal{M}_{HS}(\rho) \;, 
\end{align}
which will be used in the next section to obtain a test for $\mathcal{M}_{\Tr}(\rho)$  starting from a test for $\mathcal{M}_{HS}(\rho)$.}}

{
	If the state $\sigma$ is close to having rank $k$, in the sense that the sum of its largest $k$ eigenvalues is larger than $1-\eta$, then the following inequality (proven in section 5.4 of \cite{Badescu2019}) holds
\begin{eqnarray}
D_{\Tr}(\rho, \sigma) \leq \frac{\sqrt{k}}{c} D_{HS}(\rho, \sigma) + \eta \; ,
\end{eqnarray}
with $c = 2 - \sqrt{2}$.
Therefore, in the special case in which the average state $\bar{\rho}$ is $\eta$-close to having rank $k$, the inequality~(\ref{basic_ineq}) can be improved by
\begin{align}\label{basic_ineq_rangok}
\mathcal{M}_{\Tr}(\rho) &=\sum_{i=1}^N p_i D_{\Tr}(\rho_i, \bar\rho)\leq \sum_{i=1}^N p_i \left( \frac{1}{c}\sqrt{kD^2_{HS}(\rho_i, \bar\rho)} + \eta \right) \nonumber \\
&=  \frac{1}{c} \sum_{i=1}^N p_i \sqrt{k D^2_{HS}(\rho_i, \bar\rho)} + \eta
=
\frac{1}{c} \sum_{i=1}^N \sqrt{p_i} \sqrt{p_i k D^2_{HS}(\rho_i, \bar\rho)} + \eta \nonumber \\
&\leq \frac{1}{c}\sqrt{ \sum_{i=1}^N p_i} \sqrt{\sum_{i=1}^N p_i kD^2_{HS}(\rho_i, \bar\rho)} + \eta
= \frac{\sqrt{k}}{c\sqrt{2}}\mathcal{M}_{HS}(\rho) + \eta \;. 
\end{align}
  }

{{In our analysis we will also}} need the following divergences for classical distributions $p$,$q$: the \textit{chi-squared} divergence, defined as $\textrm{d}_{\chi^2}(p||q):=\sum_{i}\frac{(p_i-q_i)^2}{q_i}$; the \textit{Kullback-Leibler} divergence, defined as $\textrm{d}_{KL}(p||q):=\sum_{i}p_i\log_2\frac{p_i}{q_i}$; and the total variation distance, defined as $\textrm{d}_{TV}(p||q):=\frac{1}{2}\sum_{i}|p_i-q_i|$, which corresponds to the trace distance between states which are diagonal in the same basis~\cite{Cover2005a,Sason2016}.
From the definition of Kullback-Leibler divergence, it follows that it is additive, i.e.
\begin{eqnarray}
\label{additivity_KL}
\textrm{d}_{KL}\left(\prod_{j=1}^N p^{(j)}|| \prod_{j=1}^N  q^{(j)} \right) = \sum_{j=1}^N \textrm{d}_{KL}(p^{(j)}||q^{(j)}) \; .
\end{eqnarray}
{{We remind also that the total variation distance is related to the Kullback-Leibler divergence by Pinsker's inequality:
\begin{eqnarray}
\label{tv_vs_kl}
\textrm{d}_{TV}(p, q) \leq \sqrt{\frac{1}{2} \textrm{d}_{KL}(p||q)} \; ,
\end{eqnarray}
and that the Kullback-Leibler can  be bounded in terms of the chi-squared divergence, as:}}
\begin{eqnarray}
\label{kl_vs_chisq}
\textrm{d}_{KL}(p,q) \leq \ln\left[ 1 + \textrm{d}_{\chi^2}(p, q) \right] \; .
\end{eqnarray}

\subsection{Schur-Weyl duality}\label{rep}
In this section we review some key facts in group representation theory that are useful to discuss properties of i.i.d. quantum states. 
Consider  the state space of $l$, $d$-dimensional systems, 
$\mathcal{H}_d^{\otimes l}$. This space carries the action of two different groups;  the special unitary group of $d\times d$ complex matrices, $\mathrm{SU}(d)$, 
and the permutation group of $l$ objects, $S_{l}$.  Specifically, the groups $\mathrm{SU}(d)$ and $S_{l}$ act on a basis $\{\ket{i_1} \otimes \ket{i_2} \otimes ... \otimes \ket{i_{l}}\}_{i_1,i_2....,i_{l}}$ of $\mathcal{H}_d^{\otimes l}$ via unitary 
representations $u_{l}:\mathrm{SU}(d)\to\mathrm{U}(\mathcal{H}_d^{\otimes l})$, and $s_{l}:S_{l}\to\mathrm{U}(\mathcal{H}_d^{\otimes l})$ as follows
\begin{align}\nonumber
u_{l}(U)\ket{i_1} \otimes \ket{i_2} \otimes ... \otimes \ket{i_{l}}&=U^{\otimes l}\ket{i_1} \otimes \ket{i_2} \otimes ... \otimes \ket{i_{l}}\nonumber\\&=U\ket{i_1} \otimes U\ket{i_2} \otimes ... \otimes U\ket{i_{l}},\quad \forall U\in\mathrm{SU}(d)\\
s_{l}(\tau)\ket{i_1} \otimes \ket{i_2} \otimes ... \otimes \ket{i_{l}}&=\ket{\tau^{-1}({i_{1}}) }\otimes \ket{\tau^{-1}({i_2})} \otimes ... \otimes \ket{\tau^{-1}({i_{l}})},\forall\tau\in S_{l}\nonumber.
\label{supp:groupactions}
\end{align}
Observe that $[U^{\otimes l},s_{l}(\tau)]=0,\; \forall U\in\mathrm{SU}(d),\,\mathrm{and}\, \forall\tau\in S_{l}$. 
{ Let  $Y_{l,d}$ denote be the set of integer partitions of $l$ in at most $d$ parts written in decreasing order}, pictorially represented by Young diagrams, where $l$ boxes are arranged into at most $d$ rows. $\lambda\in Y_{l,d}$ can then also be written as a vector $\lambda=(\lambda_1,\lambda_2,..., \lambda_d)$ with $\lambda_1\geq\lambda_2\geq...\geq\lambda_d$.
Schur-Weyl duality~\cite{Hayashi2016b, Hayashi2017} states that the  
total state space $\mathcal{H}_d^{\otimes l}$ can be decomposed as
\begin{equation}
\mathcal{H}_d^{\otimes l}\cong\bigoplus_{\lambda\in Y_{l,d}} \,\mathcal U^{(\lambda)}(\mathrm{SU}(d))\otimes\mathcal V^{(\lambda)}(S_{l}),
\label{supp:SchurWeyl}
\end{equation}
where the unitary irreducible representation (irrep)  $u^{(\lambda)}$  of $\mathrm{SU}(d)$ acts non trivially on the factor $\mathcal U^{(\lambda)}(\mathrm{SU}(d))$ of dimension $\chi_\lambda$ and the irrep  $s^{(\lambda)}$ of $S_{l}$ acts non trivially on the factor $\mathcal V^{(\lambda)}(S_{l})$ of 
dimension $\omega_\lambda$.  
The use of the congruence sign in Eq.~\eqref{supp:SchurWeyl}  indicates that this block decomposition is accomplished by a unitary transformation; in the case considered here this unitary is the Schur transform~\cite{Bacon2006,Harrow2005,Krovi2019}.

A state $\rho^{\otimes l}\in\mathcal{D}(\mathcal{H}_d^{\otimes l})$ commutes with $s_l(\sigma)$ for any $\sigma$. By Schur's lemma, $\rho^{\otimes l}$ can be decomposed in block diagonal form according to the isomorphism in Eq.~(\ref{supp:SchurWeyl}).
\begin{equation}\label{symstate}
\rho^{\otimes l}=\sum_{\lambda\in Y_{l,d} } {\SWdens{l}{\rho}}(\lambda) \rho_\lambda \otimes \frac{\mathbf 1_\lambda}{\omega_\lambda},
\end{equation}
where ${\SWdens{l}{\rho}}(\lambda)$ is a probability distribution over the Young diagrams, which depends only on { the number of copies $l$ and on} the spectrum of $\rho$, and {$\rho_{\lambda}$ are $\chi_{\lambda}$-dimensional states}. 
Applying $u_{l}(U)$ with $U$ extracted from the Haar measure of $\mathrm{SU}(d)$ gives
\begin{equation}
\mathcal{G}_{\mathrm{SU}(d)}^l[\rho]:=\int_{U\in \mathrm{SU}(d)}\, \mathrm{d} U\, U^{\otimes l}\, \rho^{\otimes l}\, U^{\dagger\,\otimes l}=\sum_{\lambda\in Y_{l,d}} {\SWdens{l}{\rho}}(\lambda)  \frac{\mathbf 1_\lambda}{\chi_\lambda} \otimes \frac{\mathbf 1_\lambda}{\omega_\lambda}=\sum_{\lambda\in Y_{l,d}} {\SWdens{l}{\rho}}(\lambda)\frac{ \Pi_{\lambda} }{\chi_\lambda\omega_\lambda}, 
\label{supp:su(d)twirl}
\end{equation}
again by Schur's lemma, where we defined the orthogonal set of projectors $\{\Pi_{\lambda}\}_{\lambda\in Y_{l,d}}$. The projective measurement with these projectors is called weak Schur sampling~\cite{Harrow2005,Krovi2019}, and it can be executed with gate complexity $O(l,\log d,\log 1/\delta)$, where $\delta$ is the precision of the implementation {(that is, the maximum trace distance between pairs of states obtained applying the actual circuit implementation of the measurement and the ideal operation to the same pure state)}. Finally, for any decomposition $\mathcal{H}_d^{\otimes l}=\otimes_{i=1}^{N} \mathcal{H}_{d}^{\otimes m_i}$ { (where $\sum_{i=1}^N m_i = l$)}, one can define a family of weak Schur sampling projectors for each factor, $\{\Pi_{\lambda}^{(i)}\}_{\lambda\in Y_{m_i,d}}$. Since the elements of $\{\Pi_{\lambda}\}_{\lambda\in Y_{l,d}}$ commute with local permutations, they commute with the projectors $\{\otimes_{i=1}^{N}\Pi_{\lambda_i}^{(i)}\}_{\lambda_i\in Y_{m_i,d}}$. { Indeed, we can decompose  $\mathcal{H}_d^{\otimes l}$ according to irreducible representations of $S_{m_1}\times S_{m_2}\times...\times S_{m_N}$; irreducible representations are labeled by $(\lambda_1,...,\lambda_N)$,  $\lambda_i\in Y_{m_i,d}$, appear in general with multiplicity, and the projector on all the irreducible components with label $(\lambda_1,...,\lambda_N)$ is $\otimes_{i=1}^{N}\Pi_{\lambda_i}^{(i)}$. By Schur's lemma, $\{\Pi_{\lambda}\}_{\lambda\in Y_{l,d}}$ should be block diagonal according to the decomposition given by $\{\otimes_{i=1}^{N}\Pi_{\lambda_i}^{(i)}\}_{\lambda_i\in Y_{m_i,d}}$.} Therefore local and global weak Schur sampling can be done with a unique projective measurement, and the probabilities of the outcomes are the same if the two projective measurements are executed in any order. Therefore, this nested weak Schur sampling is also efficient, and it will give an implementation of the measurement required by the test we study in this paper.

\section{Upper bound on the sample complexity}\label{upbound}

In order to prove Theorem~\ref{theoremtr} here we 
show a stronger version of such statement, i.e.

\begin{theorem}\label{theoremhs}
Given access to $O(\frac{\sqrt{N}}{\delta})$ samples of the state $\rho$ of Eq.~(\ref{stato_sampling}), for $\delta >0$  there is an algorithm which can distinguish with high probability whether $\mathcal M_{HS}^2(\rho)\leq0.99 \delta$ or $\mathcal M_{HS}^2(\rho)> \delta$.
\end{theorem}
The connection with Theorem~\ref{theoremtr} follows by the relations 
between the functionals $\mathcal M_{HS}(\rho)$ and $\mathcal M_{\mathrm{Tr}}(\rho)$ discussed in Sec.~\ref{sec:dist}. Specifically we note
 that $\mathcal M_{\mathrm{Tr}}(\rho)=0$ (case $\mathsf{A}$) implies  $\mathcal M_{HS}(\rho)=0$, while  having $\mathcal M_{\mathrm{Tr}}(\rho)>\epsilon$ (a constraint that holds 
 in Case  $\mathsf{B}$) implies
 $\mathcal M_{HS}^2(\rho)>\frac{8\epsilon^2}{d}$ by Eq. (\ref{basic_ineq}). 
Therefore a test satisfying  the requests of Theorem~\ref{theoremtr}  
can be obtained by taking 
 the algorithm identified by Theorem~\ref{theoremhs} with $\delta =\frac{8\epsilon^2}{d}$. [Incidentally we stress
  that the test can be performed by a two outcome POVMs $\{E_0^{(M)},E_1^{(M)}\}$ when the number of copies of $\rho$ is $M$ (for any $M\geq 0$), obtained as projectors on the eigenvectors of the observable  $\mathcal D_{M}$,
  defined in the following, with eigenvalues respectively larger or lower than a threshold; therefore, it is of the class of test on which we can apply Proposition \ref{Poissonpropo}].
  
  {
In a complete analogous way, Theorem~\ref{theoremtr_rangok} follows by calling the algorithm of Theorem~\ref{theoremhs} with $\delta = \frac{16(2-\sqrt{2})^2\epsilon^2}{k}$, and using the inequality~(\ref{basic_ineq_rangok}). 
}

The reminder of the section is hence devoted to the prove  
Theorem~\ref{theoremhs}.

\subsection{Building the estimator for $\mathcal{M}_{HS}^2$}\label{subsec31}

To prove  Theorem~\ref{theoremhs} we construct 
 an unbiased estimator for $\mathcal{M}_{HS}^2$, generalizing the estimator of $D^2_{HS}(\rho,\sigma)$ discussed in~\cite{Badescu2019}. 
We start 
 noticing that   via permutations that operate on the quantum registers conditioned on measurements performed on the classical registers, the density matrix $\rho^{\otimes M}$ describing $M$ sampling of the state $\rho$, can be 
cast in the following  equivalent form
 \begin{equation}
 \label{stato_ordinato}
 \rho^{(M)}:=\sum_{\vec m\in {\cal P}_M} \mathsf{M}(\vec m)_{\vec p, M} \ketbra{\vec m}{\vec m} 
  \otimes \rho^{\vec m}.
 \end{equation}
In this expression  the summation runs over all vectors  $\vec{m} = (m_1, m_2, \cdots, m_N)$ formed by integers {that satisfy $m_1+m_2+\cdots +m_N=M$}; while $\mathsf{M}(\vec m)_{\vec p, M}$ is the multinomial distribution with $M$ extractions and probabilities $\vec p=(p_1, p_2, \cdots, p_N)$, i.e. 
\begin{eqnarray}
\mathsf{M}(\vec m)_{\vec p, M} := \frac{M!}{m_1!...m_N!} p_1^{m_1} p_2^{m_2} \cdots p_N^{m_N}   \;; \end{eqnarray} 
the vectors $\ket{\vec{m}}= \ket{m_1, m_2,  \cdots, m_N}$ form an orthonormal set for the classical
registers of the model; while  finally 
\begin{eqnarray} \rho^{\vec m}:=\rho_{1}^{\otimes m_1}\otimes\rho_{2}^{\otimes m_2} \otimes...\otimes\rho_{N}^{\otimes m_N}\;, \label{newversione} 
\end{eqnarray}
is a state 
of the quantum registers with $m_i$ elements initialized into $\rho_i$, which formally operates 
on a Hilbert space with tensor product structure $\otimes_{i=1}^{N} \mathcal{H}_i$, with $\mathcal H_i= (\mathbb C^{d})^{\otimes m_i}$, with $m_i=0,...,M$.
Exploiting the representation of  Eq.~(\ref{stato_ordinato}) we then introduce the 
 observable 
 \begin{align}\label{DEFDM} 
 \mathcal D_{M}&: =\sum_{\substack{\vec{m} \in {\cal P}_M}}\ketbra{\vec{m}}{\vec{m}} \otimes  \mathcal D^{\vec m,M},
 \end{align}
 with 
 \begin{equation} \label{ddfd} 
 \mathcal D^{\vec m,M}:= \sum_{i\neq j}\mathcal D_{ij}^{m_i,m_j,M},
 \end{equation}
 and
 \begin{equation}\label{defDij} 
 \mathcal D_{ij}^{m_i,m_j,M}:=\frac{m_i(m_i-1)}{\mu^2 p_i}p_j\mathcal O_{ii}^{m_i,m_i}+\frac{m_j(m_j-1)}{\mu^2 p_j}p_i\mathcal O_{jj}^{m_j,m_j}-2\frac{m_i m_j}{\mu^2}\mathcal O_{ij}^{m_i,m_j}.
 \end{equation}
 In the above expression $\mu>0$ is a free parameter that will be fixed later on. The
  operators $\mathcal O_{ij}^{m_i,m_j}$ are defined to be the average of all possible different transpositions ${S\in\,S^{m_i,m_j}_{ij}}$ between two {local copies of $\mathbb{C}^{d}$} in the spaces $\mathcal H_i$ and $\mathcal H_j$, with $i$ and $j$ possibly equal, i.e. 
\begin{equation}
\mathcal O_{ij}^{m_i,m_j}:=\frac{1}{|{S^{m_i,m_j}_{ij}}|}\sum_{S\in {S^{m_i,m_j}_{ij}}} S\;. 
\end{equation} 
Since each transposition is Hermitian, $\mathcal O_{ij}^{m_i,m_j}$ is Hermitian too. {Note that $|S^{m_i,m_j}_{ij}|=m_i m_j$ when $i\neq j$, while $|S^{m_i,m_i}_{ii}|=m_i(m_i-1)/2$.}

The expectation values of $\mathcal D_{M}$ on $\rho^{(M)}$ can be formally computed by 
exploiting the relations
{\begin{eqnarray} \Tr[\mathcal O_{ii}^{m_i,m_i}\rho^{\vec m}] = 
 \Tr[\mathcal O_{ii}^{m_i,m_i}\rho_i^{\otimes m_i}]=\Tr[\rho_i^2]\;,\end{eqnarray}
 where the first identity follows from the fact that $\mathcal O_{ii}^{m_i,m_i}$
 acts nontrivially only on registers containing copies of $\rho_i$, and}
\begin{eqnarray} \Tr[\mathcal O_{ij}^{m_i,m_j}\rho^{\vec m}] = 
 \Tr[\mathcal O_{ij}^{m_i,m_j}\rho_i^{\otimes m_i}\otimes\rho_j^{\otimes m_j}]=\Tr[\rho_i\rho_j]\;,\end{eqnarray}
 where the first identity follows from the fact that $\mathcal O_{ij}^{m_i,m_j}$
 acts not trivially only on registers containing copies of $\rho_i$ and $\rho_j$.
 Accordingly  for $i\neq j$ we have 
 \begin{equation}
 \Tr[\mathcal D_{ij}^{m_i,m_j,M} \rho^{\vec m}]=\frac{m_i(m_i-1)}{\mu^2 p_i}p_j\Tr[\rho_i^2]+\frac{m_j(m_j-1)}{\mu^2 p_j}p_i\Tr[\rho_j^2]-2\frac{m_i m_j}{\mu^2}\Tr[\rho_i\rho_j] \;, 
 \end{equation}
 which leads to 
 \begin{equation} \Tr[\mathcal D_{M}\rho^{(M)}] = \sum_{\substack{\vec{m} \in {\cal P}_M}}
\mathsf{M}(\vec m)_{\vec p, M}  \sum_{i\neq j}  \left(
\tfrac{m_i(m_i-1)}{\mu^2 p_i}p_j\Tr[\rho_i^2]+\tfrac{m_j(m_j-1)}{\mu^2 p_j}p_i\Tr[\rho_j^2]-2\tfrac{m_i m_j}{\mu^2}\Tr[\rho_i\rho_j]\right)\;.  \label{IMPO0} 
\end{equation}
To simplify the analysis of the performance of a test based on $\mathcal D_{M}$ we can invoke the equivalence of Proposition~\ref{Poissonpropo} between the original model and its Poissonized version  where the value of $M$  (and hence the density matrix $\rho^{(M)}$ that are presented to us) is  randomly generated  
with probability $\Poi_{\mu}(M)$ 
(notice that  the mean value 
 of  
 the distribution is taken equal to parameter $\mu$ which enters  the definition~(\ref{defDij}) of $D_{ij}^{m_i,m_j,M}$). Defining $\Gamma_{M}$ the set of eigenvalues of the observables $\mathcal D_M$~(\ref{DEFDM}), 
 we then introduce  a new estimator $\mathcal D$  that produces outputs $X\in \Gamma:= \bigcup_M \Gamma_M$ with probabilities
 \begin{eqnarray} \label{impo1} 
 P_X :=\sum_{M=0}^\infty \Poi_{\mu}(M) \sum_{x\in \Gamma_M}  \delta_{x,X}    P_x^{(M)}\;, 
 \end{eqnarray} 
 where $P_x^{(M)}$ is the probability of getting the outcome $x$ from  $\mathcal D_{M}$
 when acting on $\rho^{(M)}$.
 
The following facts can then be proved: 
 \begin{proposition}[Unbiasedness] \label{Unbiasedness}
 Given 
 ${\mathbb{E}} [\mathcal D]  :=  \sum_{X\in\Gamma} X P_X $ the mean value   of the estimator $\mathcal D$ we have 
 \begin{equation}\label{unbias}
 {\mathbb{E}} [\mathcal D]   = 
 \mathcal{M}_{HS}^{2}(\rho)\;. 
 \end{equation}
 \end{proposition}
 \begin{proof} 
 From Eq.~(\ref{impo1}) and (\ref{IMPO0}) we can write 
  \begin{eqnarray}\label{unbiascalc}
 {\mathbb{E}} [\mathcal D] &=& \nonumber \sum_{M=0}^\infty \Poi_{\mu}(M)  \sum_{x\in \Gamma_M}   xP_x^{(M)}=  \sum_{M=0}^\infty \Poi_{\mu}(M) \Tr[\mathcal D_{M}\rho^{(M)}] \\ \nonumber 
 &=&  \sum_{M=0}^\infty \Poi_{\mu}(M) \sum_{\substack{\vec{m} \in {\cal P}_M}}
\mathsf{M}(\vec m)_{\vec p, M} \\ &&\times  \sum_{i\neq j}  \left(
\tfrac{m_i(m_i-1)}{\mu^2 p_i}p_j\Tr[\rho_i^2]+\tfrac{m_j(m_j-1)}{\mu^2 p_j}p_i\Tr[\rho_j^2]-2\tfrac{m_i m_j}{\mu^2}\Tr[\rho_i\rho_j]\right) \nonumber \\
 &=&  \sum_{m_1=0}^\infty \cdots \sum_{m_N=0}^\infty  \Poi_{p_1\mu}(m_1)\cdots  \Poi_{p_N\mu}(m_N)  \nonumber \\ &&\times  \sum_{i\neq j}  \left(
\tfrac{m_i(m_i-1)}{\mu^2 p_i}p_j\Tr[\rho_i^2]+\tfrac{m_j(m_j-1)}{\mu^2 p_j}p_i\Tr[\rho_j^2]-2\tfrac{m_i m_j}{\mu^2}\Tr[\rho_i\rho_j]\right)\;,  \end{eqnarray}
 where in the second identity we used  $\sum_{x\in \Gamma_M}   x  P_x^{(M)}= \Tr[\mathcal D_{M}\rho^{(M)}]$, while in the last identity we
exploit the fact that  under Poissonization 
  the random variables $m_i$ become independent due to the property 
\begin{eqnarray} \label{POISIDE} \sum_{M=0}^{\infty}\Poi_\mu(M) \mathsf{M}(\vec m)_{\vec p, M}=\prod_{i=1}^{N} \Poi_{p_i\mu}(m_i)\;, \end{eqnarray}  
with $\Poi_{p_i\mu}(m_i)$ being a Poisson distribution of mean $p_i \mu$. 
Equation~(\ref{unbias}) then finally follows from the identities
\begin{eqnarray}
 \sum_{m_i=0}^\infty  m_i \; \Poi_{p_i\mu}(m_i)  = 
  \mu p_i \;, \qquad 
  \sum_{m_i=0}^\infty  \frac{m_i(m_i-1)}{p_i}\; \Poi_{p_i\mu}(m_i)=\mu^2 p_i\;. \end{eqnarray} 
 \end{proof}

 \begin{proposition}[Bound on the variance]\label{varianceboundprop}
The variance of the estimator $\mathcal D$, ${\mathrm{Var}} [\mathcal D] := \sum_{X\in \Gamma}  P_X (X- {\mathbb{E}} [\mathcal D] )^2$,  satisfies the inequality
 \begin{equation}\label{variance}
 {\mathrm{Var}} [\mathcal D]\leq O\left(\frac{N}{\mu^2} \right) + \frac{16 \mathcal{M}_{HS}^2(\rho)}{\mu}\;. 
 \end{equation}
 \end{proposition}
 
 \begin{proof}
See Appendix \ref{vbound}.
\end{proof}

{
With these ingredients we can prove Theorem~\ref{theoremhs}, following the proof of Lemma 2.1 of~\cite{Badescu2019}, which is an application of Chebyshev inequality. We reproduce here the reasoning. Let us put $c=\mathcal{M}_{HS}^2(\rho)$. By Chebyshev's inequality, $P(|\mathcal D-c|\geq \epsilon)\leq \frac{{\mathrm{Var}} [\mathcal D]}{\epsilon^2}$. If $c<0.99\delta$, then we have, for $C>0$ large enough and $\mu=C\frac{\sqrt{N}}{\delta}$, 
\begin{equation}
P(|\mathcal D-c|\geq0.005\delta)\leq\frac{{\mathrm{Var}} [\mathcal D]}{(0.005)^2\delta^2}\leq  \left(O(1) \frac{1}{C^2}+\frac{16}{C \sqrt{N}}\right)\delta^2\frac{1}{(0.005)^2\delta^2}\leq \frac{1}{3},
\end{equation}
therefore $\mathcal D\leq0.99\delta+0.005\delta=0.995\delta$ with high probability.
If $c\geq \delta$, then we have, for $C>0$ large enough and $\mu=C\frac{\sqrt{N}}{\delta}$,
\begin{equation}
P(|\mathcal D-c|\geq 0.005c)\leq\frac{{\mathrm{Var}} [\mathcal D]}{(0.005)^2c^2}\leq \left(O(1) \frac{1}{C^2}+\frac{16}{C \sqrt{N}}\right)c^2\frac{1}{(0.005)^2c^2}\leq  \frac{1}{3},
\end{equation}
therefore $\mathcal D\geq c-0.005c\geq 0.995\delta$ with high probability.

}


\section{Lower bound on the sample complexity}\label{Sec:lower} 

We now explain the idea for proving the lower bound on $M$ that follows from Theorem~\ref{theoremlow}. First of all we limit ourselves to even $d$, since for odd $d$ one can simply use the lower bound for $d-1$. We also choose the probability distribution $p$ to be uniform, $p_i=1/N$. The case $N=2$ is a straightforward consequence of the lower bound in~\cite{ODonnell2015a}, which gives a lower bound of $\Omega(d/\epsilon^2)$, noting that with access to $M$ copies of $\rho_{\epsilon}$ one can simulate access to $M$ copies of $\frac{1}{2} \left( \frac{I_d}{d} \otimes \ket{1}\bra{1} + \rho_\epsilon \otimes \ket{2}\bra{2} \right) $:
\begin{lemma}[Corollary 4.3 of~\cite{ODonnell2015a}] \label{lemma:lowerbound_wright}
	Let $\rho_\epsilon$ be a quantum state with $d/2$ eigenvalues equal to $\frac{1+2\epsilon}{d}$ and the other $d/2$ eigenvalues equal to $\frac{1-2\epsilon}{d}$. Then any algorithm that can discern between the states $(I_d / d)^{\otimes M}$ and $\rho^{\otimes M}_\epsilon$ with a probability greater than 2/3 must require $M \geq 0.15 d / \epsilon^2$.
\end{lemma}
This is a lower bound for any $N$ smaller than a constant, say $N<10$. 
Therefore we consider $N\geq10$ in the following. We define two sets of collections of $N$ quantum states. The first set $A$ contains only one collection, namely a collection where all the states are the maximally mixed states. Clearly, the only element of $A$ is a collection satisfying the property of case $\mathsf A$. For even $d$, the second set $B$ contains all the collections of states having $d/2$ eigenvalues equal to $\frac{1+8\epsilon}{d}$ and $d/2$ eigenvalues equal to $\frac{1-8\epsilon}{d}$. This means that all the states in a collection of $B$ can be written as $U_i\rho_0U_i^{\dagger}$ for $\rho_0$ with the prescribed spectrum and $U_i$ arbitrary. If each $U_i$ is drawn independently according to the Haar measure of $\mathrm{SU}(d)$,  we show that the elements of $B$ satisfy property $\mathsf B$ with probability larger than a constant. We also show an upper bound on the trace distance between $\rho_A$ and $\rho_B$, being respectively $M$ samples for a collection of all maximally mixed states and the average input of $M$ samples for collections in $B$. Explicitly, we have
\begin{equation}
\rho_A=\left(\frac{1}{N}\sum_{i=1}^{N}\ketbra{i}\otimes \frac {I}{d}\right)^{\otimes M},
\end{equation}

\begin{equation}
\rho_B=\int_{U_1,...,U_N\in \mathrm{SU}(d)}\mathrm d U_1....\mathrm d U_N\left(\frac{1}{N}\sum_{i=1}^{N}\ketbra{i}\otimes U_i\rho_0U_i^{\dagger}\right)^{\otimes M}.
\end{equation}

 If a test capable of distinguishing with high probability {between case $\mathsf A$ and case $\mathsf B$} exists, then it can be used to distinguish between $\rho_A$ and $\rho_B$. Since the probability of success in the latter task has to be lower than what we obtain from the bound on the trace distance, we obtain a lower bound on the sample complexity.

\begin{lemma}
	\label{lemma:statomedio}
Let $\{\rho_i\}_{i,...,N}$ be a collection of states such that $\frac{1}{N}\sum_{i=1}^{N}||\rho_i-\bar \rho||_1>4\epsilon$. 

Then $\frac{1}{N}\sum_{i=1}^{N}||\rho_i-\sigma||_1>{2}\epsilon$ for any $\sigma$. 
\end{lemma}
\begin{proof}
Suppose that we have $\frac{1}{N}\sum_{i=1}^{N}||\rho_i-\sigma||_1\leq2\epsilon$ for some $\sigma$. By monotonicity of the trace distance,  $||\bar\rho-\sigma||_1\leq2\epsilon$. Then
\begin{equation}
\frac{1}{N}\sum_{i=1}^{N}||\rho_i-\bar \rho||_1=\frac{1}{N}\sum_{i=1}^{N}||\rho_i-\sigma+\sigma-\bar \rho||_1\leq\frac{1}{N}\sum_{i=1}^{N}||\rho_i-\sigma||_1+||\sigma-\bar \rho||_1\leq 4\epsilon
\end{equation}
which is a contradiction.
\end{proof}

\begin{lemma}
	\label{lemma:abbastanzalontani}
For $N>10$, let $\{U_i\rho_0U_i^{\dagger}\}_{i,...,N}$ be a collection of states in $B$ and $\rho$ as in Eq. (\ref{stato_sampling}), with $p_i=1/N$. If each $U_i$ is drawn independently according to the Haar measure of $\mathrm{SU}(d)$, the probability of having $\mathcal{M}_{\Tr}(\rho) \geq {2}\epsilon$ is at least
\begin{eqnarray}
\underset{U_1, \dots, U_N \sim \mathbf{U}(d)}{\mathrm{P}} \left( \mathcal{M}_{\Tr}(\rho) >{2}\epsilon\right) \geq \frac{11}{15} \; .
\label{abbastanzalontani_tesi}
\end{eqnarray}
\end{lemma}

\begin{proof}
We denote ${\ket{k}}_{k=1,...,d}$ a basis of eigenvectors of $\rho_0$, such that 
\begin{equation}
\bra{k}\rho_0\ket{k}=\frac{1+(-1)^k8\epsilon}{d} \; ,
\label{autovalori_rho0}
\end{equation}
 and define
\begin{eqnarray}
\Theta := \sum_{k=1}^d (-1)^k \ketbra{k}{k} \, ,
\end{eqnarray}
We can write
\begin{align}
{2}\mathcal{M}_{\Tr}(\rho) &= \frac{1}{N} \sum_{i=1}^N \| \rho_i - \bar\rho \|_1 =
\frac{1}{N}\sum_{i=1}^N \left\| \rho_i - \frac{1}{N} \sum_{j=1}^N U_j \rho_0 U^\dagger_j \right\|_1  = 
\frac{1}{N}\sum_{i=1}^N \left\| U_i \rho_0 U^\dagger_i - \frac{1}{N} \sum_{j=1}^N U_j \rho_0 U^\dagger_j \right\|_1 
\nonumber 	\\ &=
\frac{1}{N}\sum_{i=1}^N \left\| \rho_0  - \frac{1}{N} \sum_{j=1}^N U^\dagger_i U_j \rho_0 U^\dagger_j U_i \right\|_1 
\geq
\frac{1}{N}\sum_{i=1}^N \sum_{k=1}^d \left\lvert \braket{k|\rho_0  - \frac{1}{N} \sum_{j=1}^N U^\dagger_i U_j \rho_0 U^\dagger_j U_i|k} \right\rvert \; .
\label{Stima_M_1}
\end{align}
{
We can observe now that, from~(\ref{autovalori_rho0}) it follows that $\frac{1\pm8\epsilon}{d}$ are maximum/minimum eigenvalues of $\rho_0$, so that
\begin{eqnarray}
\bra{k}\rho_0\ket{k} \mbox{($k$ odd)} =\frac{1-8\epsilon}{d} \leq \braket{k|U^\dagger_i U_j \rho_0 U^\dagger_j U_i|k} \leq \frac{1+8\epsilon}{d} = \bra{k}\rho_0\ket{k} \mbox{($k$ even)} \; ;
\end{eqnarray}
and therefore
\begin{eqnarray}
\sum_{k=1}^d \left\lvert \braket{k|\rho_0  - \frac{1}{N} \sum_{j=1}^N U^\dagger_i U_j \rho_0 U^\dagger_j U_i|k} \right\rvert = (-1)^k \left( \braket{k|\rho_0|k}  - \braket{k|\frac{1}{N} \sum_{j=1}^N U^\dagger_i U_j \rho_0 U^\dagger_j U_i|k} \right)
\label{chiarimento}
\end{eqnarray}
Replacing~(\ref{chiarimento}) into~(\ref{Stima_M_1}) we have
}
\begin{align}
2\mathcal{M}_{\Tr}(\rho) &\geq
\frac{1}{N}\sum_{i=1}^N \sum_{k=1}^d (-1)^k \left( \braket{k|\rho_0|k}  - \braket{k|\frac{1}{N} \sum_{j=1}^N U^\dagger_i U_j \rho_0 U^\dagger_j U_i|k} \right)
\nonumber 	\\& =
8\epsilon - \frac{1}{N^2} \sum_{i=1}^N \sum_{k=1}^d (-1)^k \sum_{j=1}^N \braket{k| U^\dagger_i U_j \rho_0 U^\dagger_j U_i|k} 
\nonumber 	\\ &=
8\epsilon - \frac{1}{N^2} \sum_{i=1}^N \sum_{j=1}^N \sum_{k=1}^d (-1)^k  \braket{k| U^\dagger_i U_j \rho_0 U^\dagger_j U_i|k} 
\nonumber 	\\ &=
8\epsilon - \frac{1}{N^2} \sum_{i=1}^N \sum_{j=1}^N \Tr\left[\hat\Theta U^\dagger_i U_j \rho_0 U^\dagger_j U_i \right]
\label{stima_M}
\end{align}
{
Now observe that $\rho_0=\frac{1}{d}(I+8\epsilon\Theta)$. Therefore 
\begin{align}
\sum_{i=1}^N \sum_{j=1}^N \Tr\left[\hat\Theta U^\dagger_i U_j \rho_0 U^\dagger_j U_i \right]&=\frac{1}{d}\sum_{i=1}^N \sum_{j=1}^N \Tr\left[\hat\Theta U^\dagger_i U_j U^\dagger_j U_i \right]+\frac{8\epsilon}{d}\sum_{i=1}^N \sum_{j=1}^N \Tr\left[\hat\Theta U^\dagger_i U_j \hat \Theta U^\dagger_j U_i \right]\nonumber\\&=\frac{8\epsilon}{d}\sum_{i=1}^N \Tr\left[\left(\sum_{i=1}^N U_i \hat\Theta U^\dagger_i \right) \left(\sum_{j=1}^N  U_j \hat{\Theta} U^\dagger_j  \right)\right]\geq 0.
\end{align} 
Since the latter term of~(\ref{stima_M}) is always positive, we may use the Markov's inequality on it. }
Its expected value is:
\begin{align}
&\underset{U_1, \dots, U_N \sim \mathbf{U}(d)}{\mathbb{E}}  \left[\frac{1}{N^2} \sum_{i=1}^N \sum_{j=1}^N \Tr\left[\hat\Theta U^\dagger_i U_j \rho_0 U^\dagger_j U_i \right] \right]
= 
\frac{1}{N^2} \sum_{j=1}^N \sum_{i=1}^N \underset{U_1, \dots, U_N \sim \mathbf{U}(d)}{\mathbb{E}} \left[\Tr\left[\hat\Theta U^\dagger_i U_j \rho_0 U^\dagger_j U_i \right] \right] 
\nonumber \\ &= 
\frac{1}{N^2} \sum_{j=1}^N \sum_{i=1}^N 8\epsilon\delta_{ij}
=
8\frac{\epsilon}{N} \; . 
\end{align}
Therefore, using Markov inequality, we can write
\begin{eqnarray}
\underset{U_1, \dots, U_N \sim \mathbf{U}(d)}{\mathrm{P}} \left( \frac{1}{N^2} \sum_{i=1}^N \sum_{j=1}^N \Tr\left[\hat\Theta U^\dagger_i U_j \rho_0 U^\dagger_j U_i \right]\geq 3\epsilon\right) \leq \frac{8}{3N} 
\label{markov1}
\end{eqnarray}
Combining~(\ref{markov1}) with~(\ref{stima_M}), we have
\begin{eqnarray}
\underset{U_1, \dots, U_N \sim \mathbf{U}(d)}{\mathrm{P}} \left( \mathcal{M}_{\Tr}(\rho) > {2}\epsilon\right) \geq 1 - \frac{8}{3N}\geq \frac{11}{15},  \qquad N\geq 10
\label{markov2}
\end{eqnarray}

\end{proof}

\begin{lemma}
	\label{lemma:distanzaTV}
	\begin{equation}
	D_{\mathrm{Tr}}(\rho_A,\rho_B)\leq 16 \frac{\epsilon^2M}{d\sqrt{N}}
	\end{equation}
\end{lemma}

\begin{proof}

We have that 

\begin{equation}
D_{\mathrm{Tr}}(\rho_A,\rho_B)=E_{\vec m \sim \mathsf{M}_{\vec p, N,M}} \left[D\left(\left(\frac {I}{d}\right)^{\otimes M}, \int_{U_i\in \mathrm{SU}(d)}\mathrm d U_1....\mathrm d U_N\bigotimes_{i=1}^{N}\left(U_i\rho_0U_i^{\dagger}\right)^{\otimes m_i}\right)\right]
\end{equation}

Using Schur-Weyl duality, we can write $\rho_A$ and $\rho_B$ as

\begin{equation}
\left(\frac {I}{d}\right)^{\otimes M}=\bigotimes_{i=1}^{N}\left(\sum_{\lambda\in Y_{m_i,d}} {\SWdens{m_i}{I/d}}(\lambda) \frac{I_{d(\lambda,m_i)\times d(\lambda,m_i)}}{d(\lambda,m_i)} \right)
\end{equation}

\begin{equation}
 \int_{U_i\in \mathrm{SU}(d)}\mathrm d U_1....\mathrm d U_N\bigotimes_{i=1}^{N}\left(U_i\rho_0U_i^{\dagger}\right)^{\otimes m_i}=\bigotimes_{i=1}^{N}\left(\sum_{\lambda\in Y_{m_i,d}} {\SWdens{m_i}{\rho_0}}(\lambda) \frac{I_{d(\lambda,m_i)\times d(\lambda,m_i)}}{d(\lambda,m_i)} \right),
\end{equation}
where $Y_{m_i,d}$ is a set of Young diagrams and $\SWdens{M}{\rho}(\lambda)$ is a probability distribution over Young diagrams which depends only on the spectrum of $\rho$. 
Defining

\begin{eqnarray}
\mathfrak{D}^{\vec{m}}_{0} = \SWdens{m_1}{d} \times \cdots \SWdens{m_i}{d}\;, \qquad \mathfrak{D}^{\vec{m}}_{\epsilon} = \SWdens{m_1}{\rho_0} \times \cdots \SWdens{m_i}{\rho_0},
\label{distribuzione_id}
\end{eqnarray}

we have
\begin{equation}
D_{\mathrm{Tr}}(\rho_A,\rho_B)=E_{\vec m \sim \mathsf{M}_{\vec p, N,M}}d_{TV}(\mathfrak{D}^{\vec{m}}_{0}, \mathfrak{D}^{\vec{m}}_{\epsilon} )
\end{equation}
First of all we invoke the from~\cite{ODonnell2015a}:
\begin{eqnarray}
\label{chiquadro_1copia}
\mathrm{d}_{\chi^2}(\SWdens{n}{\rho} || \SWunif{n}{I/d} ) \leq \exp(256n^2\epsilon^4/d^2) - 1
\end{eqnarray}

Our first observation is that, when $m_i = 1$,~(\ref{chiquadro_1copia}) can be improved noticing that $d_{KL} (\SWdens{1}{\rho_i}, \SWdens{1}{I/d}) = 0$ for every possible state $\rho_i$ (since there is only one possible partition of $n=1$ - in other words, we gain no information on whether the state is mixed by measuring a single copy). This observation, together with~(\ref{chiquadro_1copia}) and~(\ref{kl_vs_chisq}), imply that
\begin{eqnarray}
\mathrm{d}_{KL}(\SWdens{m_i}{\rho} || \SWunif{m_i}{I/d} )\leq 256 \frac{1_{m_i > 1} \cdot m_i^2 \epsilon^4}{d^2} \; .
\label{bound_kl}
\end{eqnarray}

Using~(\ref{additivity_KL}) and~(\ref{bound_kl}) we can write

\begin{align}
D_{\mathrm{Tr}}(\rho_A,\rho_B)&=E_{\vec m \sim \mathsf{M}_{\vec p, N,M}}d_{TV}(\mathfrak{D}^{\vec{m}}_{0}, \mathfrak{D}^{\vec{m}}_{\epsilon} )\leq E_{\vec m \sim \mathsf{M}_{\vec p, N,M}}\sqrt{\frac{1}{2}d_{KL}(\mathfrak{D}^{\vec{m}}_{0}, \mathfrak{D}^{\vec{m}}_{\epsilon} })\nonumber\\
&=E_{\vec m \sim \mathsf{M}_{\vec p, N,M}}\sqrt{\frac{1}{2}\sum_{i=1}^{N}\mathrm{d}_{KL}(\SWdens{m_i}{\rho} || \SWunif{m_i}{I/d} )} \leq E_{\vec m \sim \mathsf{M}_{\vec p, N,M}}\sqrt{\frac{1}{2}\sum_{i=1}^{N} 256\frac{1_{m_i > 1} \cdot m_i^2 \epsilon^4}{d^2}}\nonumber\\
&\leq\sqrt{E_{\vec m \sim \mathsf{M}_{\vec p, N,M}}\frac{1}{2}\sum_{i=1}^{N} 256\frac{1_{m_i > 1} \cdot m_i^2 \epsilon^4}{d^2}}\leq\sqrt{E_{\vec m \sim \mathsf{M}_{\vec p, N,M}}\sum_{i=1}^{N} 256m_i(m_i-1)\frac{\epsilon^4}{d^2}}\nonumber\\
&\leq16 \frac{\epsilon^2M}{d\sqrt{N}},
\end{align}
where the first inequality is from Pinsker's inequality, the second equality is the additivity of the Kullback-Leibler divergence, the third inequality is from concavity of the square root.
\end{proof}

It is now immediate to prove Theorem \ref{theoremlow}
\begin{proof}[Proof of Theorem \ref{theoremlow}]
If an algorithm as in Theorem~\ref{theoremlow} exists, one can use it to try to discriminate between $\rho_A$ and $\rho_B$. By also invoking the Holevo-Helstrom bound Eq.~(\ref{HH}), the probability of success has to satisfy
\begin{equation}
\frac{1}{2}\left(1+16 \frac{\epsilon^2M}{d\sqrt{N}}\right)
\geq p_{succ}\geq \frac{1}{2}\left(\frac{11}{15}+1\right)\frac{2}{3} \; .
\end{equation}
Therefore
\begin{equation}
M\geq 4\cdot 10^{-3} \frac{\sqrt{N}d}{\epsilon^2}.
\end{equation}
\end{proof}

\section{Implementation of the optimal measurement}

The measurement of the test defined in Section~\ref{upbound} to prove Theorem~\ref{theoremtr} can be implemented on a quantum computer with gate complexity $O(M, \log d, \log 1/\delta)$, where $\delta$ is the precision of the implementation, because it can be realized with a sequence of weak Schur sampling measurements. This was already shown for the observable of~\cite{Badescu2019} for $N=2$ and it can be easily be shown to be true in the general case too. Indeed, in~\cite{Badescu2019} it is shown that $\mathcal O_{ii}^{m_i,m_i}$ can be written as

\begin{equation}
\mathcal O_{ii}^{m_i,m_i}=\sum_{\lambda\in Y_{m_i,d}} \mathsf{TN(\lambda)}\Pi^{(i)}_{\lambda},
\end{equation}

where $Y_{m_i,d}$ are Young diagrams, $\Pi_{\lambda}$ a complete set of orthogonal projectors and $\mathsf{TN(\lambda)}=\frac{1}{n(n-1)}\sum_{i=1}^{d}((\lambda_i-i+1/2)^2-(-i+1/2)^2)$. We now define $\mathcal O$ to be the average of all transposition on $\mathcal H_{d}^{\otimes M}$, for which we have:

\begin{equation}
\mathcal O=\sum_{\lambda\in Y_{M,d}} \mathsf{TN(\lambda)}\Pi_{\lambda}.
\end{equation}

Using that 

\begin{equation}
\frac{M(M-1)}{2}\mathcal O= \frac{1}{2}\sum_{i\neq j} m_i m_j \mathcal{O}_{ij}^{m_i,m_j}+ \sum_{i=1}^{N} \frac{m_i(m_i-1)}{2} \mathcal{O}_{ii}^{m_i,m_i},
\end{equation}
we have
\begin{align}
\mathcal D^{\vec m,M}&:= \sum_{i\neq j}\mathcal D_{ij}^{m_i,m_j,M}={\sum_{i=1}^{N}\frac{2m_i(m_i-1)(1-p_i)}{\mu^2 p_i}\mathcal{O}_{ii}^{m_i,m_i}-\sum_{i\neq j}2\frac{m_i m_j}{\mu^2}\mathcal O_{ij}^{m_i,m_j}.}
\nonumber\\
&=\sum_{i=1}^{N}\frac{2m_i(m_i-1)}{\mu^2 p_i}\mathcal{O}_{ii}^{m_i,m_i}-\frac{2M(M-1)}{\mu^2}\mathcal O.
\end{align}

Since $[\Pi_{\lambda},\otimes_{i=1}^{N}\Pi_{\lambda_i}^{(i)}]=0$, the measurement can be implemented efficiently by nested weak Schur sampling.

\section{Conclusions and remarks}
We have established the sample complexity of testing identity of collections of quantum states in the sampling model, with a test that can be also implemented efficiently in terms of gate complexity. Note that for this problem one could have used the independence tester of~\cite{yu2019quantum}, based on the identity test of~\cite{Badescu2019}, since if the state in the collection are equal the input of our problem in Eq.~(\ref{stato_sampling}) is a product state, and far from it otherwise. However, the guaranteed sample complexity in this case would have been $O(Nd/\epsilon^2)$, and to get $\sqrt{N}d/\epsilon^2$ we need to make use of the fact that the state in Eq.~(\ref{stato_sampling}) is a classical-quantum state and that we know the classical marginal. This is a state of zero discord~\cite{Henderson2001,Ollivier2001,Adesso2016}, and one could ask how the sample complexity differ if the discord is not zero, for example if the states $\ket{i}$ are not orthogonal. This could be seen as an example of quantum inference problem with quantum flags, proved useful in other contexts, e.g. the evaluation of quantum capacities~\cite{Smith2008b,Leditzky2018b,Fanizza2020,Kianvash2020,Wang2021,fanizza2021estimating}. More generally, an interesting problem would be to study the sample complexity of independence testing with constraints on the structure of the state, with a rich variety of scenarios possible. 

\section{Acknowledgment}
M. F. thanks A. Montanaro for suggesting the problem, M. Rosati, M. Skotiniotis and J. Calsamiglia for many discussions about distance estimation, and M. Christandl, M. Hayashi and A. Winter for helpful comments. The authors acknowledge support by MIUR via PRIN 2017 (Progetto di Ricerca di Interesse Nazionale): project QUSHIP (2017SRNBRK). MF is supported by a Juan de la Cierva Formaci\`on fellowship (project FJC2021-047404-I), with funding from MCIN/AEI/10.13039/501100011033 and European Union NextGenerationEU/PRTR, and by 
 Spanish Agencia Estatal de Investigación, project PID2019-107609GB-I00/AEI/10.13039/501100011033, by the European Union Regional Development Fund within the ERDF Operational Program of Catalunya (project QuantumCat, ref. 001-P-001644), and by European Space Agency, project ESA/ESTEC 2021-01250-ESA.
{\raggedright
        \printbibliography
        
}

\begin{appendix}
\section{Equivalence of sampling model and Poissonized model}\label{Poissonapp}
The equivalence of the Poisson model with the original one can be formalised in the following propositions.

\begin{proposition}\label{Poissonpropo}Suppose that given access to $M$ copies of the state $\rho$ of Eq.~(\ref{stato_sampling}), where $M$ is extracted from a Poisson distribution with mean $\mu$, there is a test \texttt{Ptest} such that 
\begin{eqnarray} 
	\left\{ \begin{array}{l}  \label{pconstraint} 
	P( \mbox{\texttt{Ptest $\mapsto$ "accept"} }| \mbox{Case $\mathsf{A}$}  ) >  3/4 \;, \\ \\
	P( \mbox{\texttt{Ptest $\mapsto$ "accept"} }|  \mbox{Case $\mathsf{B}$} )<  1/4\;,\end{array} \right.
	\end{eqnarray}
and it can be performed by a two-outcome POVM $\{E_0^{(M)},E_1^{(M)}\}$ for each $M$.
Then, provided that $\mu$ is larger than a fixed constant, there is a test in the sampling model using $2\mu$ copies of $\rho$ satisfying
\begin{eqnarray} 
	\left\{ \begin{array}{l}  \label{constraint2} 
	P( \mbox{\texttt{test $\mapsto$ "accept"} }| \mbox{Case $\mathsf{A}$}  ) >  2/3 \;, \\ \\
	P( \mbox{\texttt{test $\mapsto$ "accept"} }|  \mbox{Case $\mathsf{B}$} )<  1/3\;. \end{array} \right.
	\end{eqnarray}
\end{proposition}
\begin{proof}
Given $2\mu$ copies of $\rho$, we construct the following test. We extract $M$ from a Poisson distribution with mean $\mu$. If $M\leq2\mu$, we perform the measurement $\{E_0^{(M)},E_1^{(M)}\}$, otherwise we declare failure.
The difference of the acceptance probabilities of \texttt{test} and \texttt{Ptest} is
\begin{eqnarray}
&&P(\texttt{Ptest} \mapsto \texttt{"accept"} )-P( \texttt{test} \mapsto\texttt{"accept"} )\nonumber\\&&\qquad =\sum_{M=0}^{2\mu}\Poi_{\mu}(M)\left(\Tr[E_0^{(M)} \rho^{\otimes M}]-\Tr[E_0^{(M)} \rho^{\otimes M}]\right)+\sum_{M=2\mu+1}^{\infty}\Poi_{\mu}(M)\left(\Tr[E_0^{(M)} \rho^{\otimes M}]-0\right)\nonumber\\
&&\qquad =\sum_{M=2\mu+1}^{\infty}\Poi_{\mu}(M)\Tr[E_0^{(M)} \rho^{\otimes M}],
\end{eqnarray}
which implies 
\begin{align} \label{questa1} 
0&\leq P(\texttt{Ptest} \mapsto \texttt{"accept"} )-P( \texttt{test} \mapsto\texttt{"accept"} )\leq \sum_{M=2\mu+1}^{\infty} \Poi_{\mu}(M)=P_{M\sim \Poi_\mu}(M>2\mu).
\end{align}
Invoking hence the Cramér-Chernoff tail bound on the Poisson distribution~\cite{Boucheron2013}, i.e.
\begin{eqnarray} P_{M\sim \Poi_\mu}(M\geq t)\leq e^{-{\mu} h(t/\mu)}\qquad h(x)=(1+x)\log (1+x)-x\;, \end{eqnarray} 
and setting {$\mu>2$}, from  Eq.~(\ref{questa1}) we then get 
\begin{align}
0&\leq P(\texttt{Ptest} \mapsto \texttt{"accept"} )-P( \texttt{test} \mapsto\texttt{"accept"} )\leq 
e^{-\mu h(2)} < 1/10\;, 
\end{align}
from which the statement of the proposition follows.
\end{proof}

{\begin{proposition}\label{Poissonpropo2}Suppose that given access to $M$ copies of the state $\rho$ of Eq.~(\ref{stato_sampling}), there is a test \texttt{Ptest} such that 
\begin{eqnarray} 
	\left\{ \begin{array}{l}  \label{pconstrainta} 
	P( \mbox{\texttt{Ptest $\mapsto$ "accept"} }| \mbox{Case $\mathsf{A}$}  ) >  3/4 \;, \\ \\
	P( \mbox{\texttt{Ptest $\mapsto$ "accept"} }|  \mbox{Case $\mathsf{B}$} )<  1/4\;,\end{array} \right.
	\end{eqnarray}
and it can be performed by a two-outcome POVM $\{E_0^{(M)},E_1^{(M)}\}$.
Then, provided that $M$ is larger than a fixed constant, there is a test in the Poissonized sampling model using $M'$ copies of $\rho$  where $M'$ is extracted from a Poisson distribution with mean $2M$, satisfying
\begin{eqnarray} 
	\left\{ \begin{array}{l}  \label{constraint2a} 
	P( \mbox{\texttt{test $\mapsto$ "accept"} }| \mbox{Case $\mathsf{A}$}  ) >  2/3 \;, \\ \\
	P( \mbox{\texttt{test $\mapsto$ "accept"} }|  \mbox{Case $\mathsf{B}$} )<  1/3\;. \end{array} \right.
	\end{eqnarray}
\end{proposition}

\begin{proof}
We have that~\cite{Boucheron2013},
\begin{eqnarray} P_{M'\sim \Poi_{2M}}(M'\leq t)\leq e^{-2M h(-t/(2M))}\qquad h(x)=(1+x)\log (1+x)-x\;.\end{eqnarray} 
Therefore, if $M>16$
\begin{eqnarray} P_{M'\sim \Poi_{2M}}(M'\leq M)\leq e^{-2M h(-1/2)}<1/10 .\end{eqnarray} 
with high probability $M'>M$ and we can use $\texttt{Ptest}$ on $M$ copies.
\end{proof}
}

\section{Proof of Proposition \ref{varianceboundprop}}\label{vbound}

As in the proof of Proposition~\ref{Unbiasedness}
 we  can invoke  Eqs.~(\ref{impo1}), (\ref{IMPO0}) and the identity $\sum_{x\in \Gamma_M}   x^2  P_x^{(M)}= \Tr[\mathcal D^2_{M}\rho^{(M)}]$ to write 
\begin{eqnarray} 
{\Var} [\mathcal D] &=&   \sum_{M=0}^\infty \Poi_{\mu}(M) \Tr[\mathcal D_{M}^2\rho^{(M)}] - {\mathbb{E}} [\mathcal D]^2 \nonumber \\ \label{IMPOIMPO1} 
&=& \sum_{M=0}^\infty \Poi_{\mu}(M) \sum_{\vec m\in {\cal P}_M} \mathsf{M}(\vec m)_{\vec p, M} \Tr[ \mathcal (D^{\vec m,M})^2 \rho^{\vec{m}}] - {\mathbb{E}} [\mathcal D]^2\;,
\end{eqnarray} 
where the last passage involves  (\ref{DEFDM}) and (\ref{stato_ordinato}). Replacing Eqs.~(\ref{stato_ordinato}), (\ref{ddfd}), and (\ref{defDij}) into 
 $\Tr[ \mathcal (D^{\vec m,M})^2 \rho^{\vec{m}}]$ reveals that such term can be written as a linear combination of
the expectation values of the operators $\mathcal O_{ij}^{m_i,m_j}\mathcal O_{kl}^{m_k,m_l}$ on $\rho^{\vec m}$   which are complicated functions of 
of the random variable $m_i$ and traces of powers of the $\rho_i$ reported in the next subsection. 
 Invoking hence~(\ref{POISIDE}) to decouple the averages over the $m_i$ we can finally write 
\begin{align}
\Var[\mathcal{D}]&=V_1+V_2\;, 
\end{align}
where setting $\Var_{\rho}[O]:=\Tr[(O-\Tr[O\rho])^2\rho]$, we defined
\begin{align}
V_1&=\underset{\substack{m_l \sim \Poi(p_l\mu) \\ l = 1, \dots, N}}{\mathbb{E}} \Var_{\rho^{\vec m}}\left[D^{\vec m,M} \right] \label{def_V1}, \\
V_2&= \underset{\substack{m_l \sim \Poi(p_l\mu) \\ l = 1, \dots, N}}{\mathbb{E}}  \left( \Tr\left[\mathcal{D}^{\vec m,M}\rho^{\vec m}\right]-\sum_{i,j }p_i p_jD^2_{HS}(\rho_i,\rho_j) \right)^2 \; , \label{def_V2}
\end{align}
 (we remind that the expression $m_l \sim \Poi(p_l\mu)$ indicates that the random variables $m_l$ are extracted from a Poisson distribution of mean $p_l\mu$).

\subsection{Bound on $V_1$}

The covariance of two observables $O,\,O'$ on a state $\rho$ is defined as 
\begin{equation}
\Cov_{\rho}[O,O']:=\Tr[(O-\Tr[O\rho])(O'-\Tr[O'\rho])\rho].
\end{equation} 
The covariances of the observables $\mathcal O_{ij}^{m_i,m_j}$ on $\rho^{\vec m}$, read:

\begin{align}
\Var_{\rho^{\vec m}}[{\mathcal{O}_{ii}^{m_i,m_i}}] 
&= \frac{2}{m_i (m_i - 1)}(1- (\Tr[\rho^2_i])^2 ) + \frac{4(m_i-2)}{m_i(m_i - 1)}(\Tr[\rho^3_i] - (\Tr[\rho^2_i])^2)\label{lem:purity-variance}\\
\Var_{\rho^{\vec m}}[\mathcal{O}_{ij}^{m_i,m_j}]
&= \frac{1}{m_im_j}
+ \frac{1 - m_i - m_j}{m_im_j} \Tr[\rho_i \rho_j]^2 \nonumber \\
&+ \frac{1}{m_j}\left(1-\frac{1}{m_i}\right) \Tr[\rho^2_i \rho_j]
+ \frac{1}{m_i}\left(1-\frac{1}{m_j}\right)  \Tr[\rho_i \rho^2_j]\qquad i\neq j\label{prop:variance-linear-fidelity}  \\
\Cov_{\rho^{\vec m}}[{\mathcal{O}_{ii}^{m_i,m_i}}, {\mathcal{O}_{ij}^{m_i,m_j}}]&= \frac{2}{m_i} \left( \Tr[\rho^2_i\rho_j] - \Tr[\rho^2_i]\Tr[\rho_i\rho_j] \right)\qquad i\neq j \label{covarianza_ii_ij}\\
\Cov_{\rho^{\vec m}}[{\mathcal{O}_{ij}^{m_i,m_j}}, {\mathcal{O}_{ik}^{m_i,m_k}}]&= \frac{\Tr[\rho_i\rho_j\rho_k]-\Tr[\rho_i\rho_j]\Tr[\rho_i\rho_k]}{m_i} \qquad i\neq j\wedge i\neq k \wedge j\neq k\label{covarianza_ij_ik}\\
\Cov_{\rho^{\vec m}}[{\mathcal{O}_{ij}^{m_i,m_j}}, {\mathcal{O}_{kl}^{m_k,m_l}}]&= 0 \qquad {i,j,k,l \,\,\,\text{all different}},
\end{align}

Replacing the above expressions
into~(\ref{def_V1}), we can rewrite it as
\begin{align}
V_1&=\underset{\substack{m_l \sim \Poi(p_l\mu) \\ l = 1, \dots, N}}{\mathbb{E}} \Var_{\rho^{\vec m}}\left[\sum_{i\neq j} \left(\frac{m_i(m_i-1)}{\mu^2p_i}{p_j}\mathcal O^{m_i,m_i}_{ii}+ \frac{m_j(m_j-1)}{\mu^2p_j}p_i\mathcal O^{m_j,m_j}_{jj}-2\frac{m_i m_j}{\mu^2}\mathcal O^{m_i,m_j}_{ij}\right)\right]\nonumber\\
&=\sum_{i} 4\underset{m_i \sim \Poi(p_i\mu)}{\mathbb{E}}\frac{m_i^2(m_i-1)^2}{\mu^4p_i^2}(1-p_i)^2\Var[{\mathcal{O}_{ii}^{m_i,m_i}}]+8\sum_{i\neq j} \underset{\substack{m_i \sim \Poi(p_i\mu) \\ m_j \sim \Poi(p_j\mu)}}{\mathbb{E}}\frac{m_i^2m_j^2}{\mu^4}\Var[{\mathcal{O}_{ij}^{m_i,m_j}}]\nonumber\\
&-16\sum_{i\neq j}  \underset{\substack{m_i \sim \Poi(p_i\mu) \\ m_j \sim \Poi(p_j\mu)}}{\mathbb{E}} \frac{m_i^2(m_i-1)m_j}{\mu^4 p_i}(1-p_i)\Cov[{\mathcal{O}_{ii}^{m_i,m_i}}, {\mathcal{O}_{ij}^{m_i,m_j}}]
\nonumber \\ &+{16}\sum_{i\neq j\neq k\neq i}\underset{\substack{m_i \sim \Poi(p_i\mu) \\ m_j \sim \Poi(p_j\mu) \\ m_k \sim \Poi(p_k\mu)}}{\mathbb{E}}\frac{m_i^2 m_j m_k}{\mu^4}\Cov[{\mathcal{O}_{ij}^{m_i,m_j}}, {\mathcal{O}_{ik}^{m_i,m_k}}] 
\label{contone1}
\end{align}

Now we proceed to evaluate separately each term of~(\ref{contone1}).

From~(\ref{lem:purity-variance}) we get
\begin{align}
&\underset{m_i \sim \Poi(p_i\mu)}{\mathbb{E}} \left[ \frac{m_i^2(m_i-1)^2}{\mu^4p_i^2}(1-p_i)^2\Var[{\mathcal{O}_{ii}^{m_i,m_i}}] \right]
\nonumber\\&=
\underset{m_i \sim \Poi(p_i\mu)}{\mathbb{E}} \left[ \frac{m_i(m_i-1)}{\mu^4p_i^2}(1-p_i)^2 [2(1- (\Tr[\rho^2_i])^2 ) + 4(m_i-2)(\Tr[\rho^3_i] - (\Tr[\rho^2_i])^2] \right] \nonumber\\
&=
{ \frac{\mu^2p^2_i}{\mu^4p_i^2}(1-p_i)^2 [2(1- (\Tr[\rho^2_i])^2 ) + 4\mu p_i(\Tr[\rho^3_i] - (\Tr[\rho^2_i])^2] } \nonumber\\
&\leq\frac{4p_i(1-p_i)^2}{\mu}(\Tr[\rho^3_i] - (\Tr[\rho^2_i])^2)+O({{1}}/\mu^2)
\label{contone2a}
\end{align}
{ where in the third line we used the fact that $\mathbb{E}[m_i(m_i-1)] = \mu^2_ip^2_i$ and $\mathbb{E}[m_i(m_i-1)(m_i-2)] = \mu^3_ip^3_i$ for a Poisson distribution with mean $\mu_ip_i$.}

Analougously, from~(\ref{prop:variance-linear-fidelity}) we have
\begin{align}
&\underset{\substack{m_i \sim \Poi(p_i\mu) \\ m_j \sim \Poi(p_j\mu)}}{\mathbb{E}}\left[\frac{m_i^2m_j^2}{\mu^4}\Var[{\mathcal{O}_{ij}^{m_i,m_j}}]\right]\nonumber\\&=\underset{\substack{m_i \sim \Poi(p_i\mu) \\ m_j \sim \Poi(p_j\mu)}}{\mathbb{E}}\left[\frac{m_im_j}{\mu^4}(1
+ (1 - m_i - m_j)\Tr[\rho_i \rho_j]^2 \nonumber + \left(m_i-1\right) \Tr[\rho^2_i \rho_j]
+ \left(m_j-1\right) \Tr[\rho_i \rho^2_j])\right]\nonumber\\
&\leq\frac{p_i p_j^2 \Tr[\rho_i\rho_j^2]+p_j p_i^2 \Tr[\rho_j\rho_i^2]-p_ip_j(p_i+p_j)\Tr[\rho_i\rho_j]^2}{\mu}+{\frac{2p_ip_j}{\mu^2}}
\label{contone2b}
\end{align}

The corresponding contribution from~(\ref{covarianza_ii_ij}) is
\begin{align}
&\underset{\substack{m_i \sim \Poi(p_i\mu) \\ m_j \sim \Poi(p_j\mu)}}{\mathbb{E}} \left[\frac{m_i^2(m_i-1)m_j}{\mu^4 p_i}(1-p_i)\Cov[{\mathcal{O}_{ii}^{m_i,m_i}}, {\mathcal{O}_{ij}^{m_i,m_j}}]\right]\nonumber\\
&=\underset{\substack{m_i \sim \Poi(p_i\mu) \\ m_j \sim \Poi(p_j\mu)}}{\mathbb{E}}\left[ \frac{m_i(m_i-1)m_j}{\mu^4 p_i}(1-p_i)2\left( \Tr[\rho^2_i\rho_j] - \Tr[\rho^2_i]\Tr[\rho_i\rho_j] \right)\right]\nonumber \\
&=\frac{(1-p_i) p_ip_j}{\mu}2\left( \Tr[\rho^2_i\rho_j] - \Tr[\rho^2_i]\Tr[\rho_i\rho_j] \right)
\label{contone2c}
\end{align}

Finally, from~(\ref{covarianza_ij_ik}) we have
\begin{align}
&\underset{\substack{m_i \sim \Poi(p_i\mu) \\ m_j \sim \Poi(p_j\mu) \\ m_k \sim \Poi(p_kM)}}{\mathbb{E}} \left[\frac{m_i^2 m_j m_k}{\mu^4}\Cov[{\mathcal{O}_{ij}^{m_i,m_j}}, {\mathcal{O}_{ik}^{m_i,m_k}}]\right]\nonumber\\&=\underset{\substack{m_i \sim \Poi(p_i\mu) \\ m_j \sim \Poi(p_j\mu) \\ m_k \sim \Poi(p_kM)}}{\mathbb{E}} \left[\frac{m_i m_j m_k}{\mu^4}\left( \Tr[\rho_i\rho_j\rho_k]-\Tr[\rho_i\rho_j]\Tr[\rho_i\rho_k]\right)\right]\nonumber \\
&=\frac{p_ip_jp_k}{\mu}(\Tr[\rho_i\rho_j\rho_k]-\Tr[\rho_i\rho_j]\Tr[\rho_i\rho_k])
\label{contone2d}
\end{align}

Inserting~(\ref{contone2a}),~(\ref{contone2b}) and~(\ref{contone2d}) into~(\ref{contone1}) we can finally write
\begin{align}
V_1&=16\sum_{i}\frac{p_i(1-p_i)^2}{\mu}(\Tr[\rho^3_i] - (\Tr[\rho^2_i])^2) \nonumber\\&+8\sum_{i\neq j} \frac{p_i p_j^2 \Tr[\rho_i\rho_j^2]+p_j p_i^2 \Tr[\rho_j\rho_i^2]-p_ip_j(p_i+p_j)\Tr[\rho_i\rho_j]^2}{\mu}\nonumber\\
&-32\sum_{i\neq j} \frac{(1-p_i) p_ip_j}{\mu}\left( \Tr[\rho^2_i\rho_j] - \Tr[\rho^2_i]\Tr[\rho_i\rho_j] \right)\nonumber\\&+{16}\sum_{i\neq j\neq k\neq i}\frac{p_ip_jp_k}{\mu}(\Tr[\rho_i\rho_j\rho_k]-\Tr[\rho_i\rho_j]\Tr[\rho_i\rho_k])+O(N/\mu^2)
\label{risultato_V1}
\end{align}

\subsection{Bound on $V_2$}

We start defining the quantities
\begin{align}
o_{ii}=\left(\frac{m_i(m_i-1)}{\mu^2 p_i}-p_i\right)\Tr[\rho_i^2], \qquad o_{ij}=\left(\frac{m_im_j}{\mu^2}-p_ip_j \right)\Tr[\rho_i\rho_j], \qquad i\neq j.
\end{align}
Noting that
\begin{equation}
\Tr\left[\mathcal{D}^{\vec m,\mu}\rho^{\vec m}\right]-\sum_{i,j}p_i p_j D^2_{HS}(\rho_i,\rho_j) =\sum_{i\neq j} \left( p_jo_{ii}+p_io_{jj}-2o_{ij}\right) \; ,
\end{equation}
we can rewrite~(\ref{def_V2}) as
\begin{align}
V_2&=\sum_{i}4(1-p_i)^2 \underset{m_i \sim \Poi(p_i\mu)}{\mathbb{E}} [o_{ii}^2]+8\sum_{i\neq j} \underset{\substack{m_i \sim \Poi(p_i\mu) \\ m_j \sim \Poi(p_j\mu)}}{\mathbb{E}}[o_{ij}^2]
\nonumber \\
&+{16}\sum_{i\neq j\neq k\neq i} \underset{\substack{m_i \sim \Poi(p_i\mu) \\ m_j \sim \Poi(p_j\mu) \\ m_k \sim \Poi(p_k\mu)}}{\mathbb{E}}[o_{ij}o_{ik}]-16\sum_{i\neq j}(1-p_i)\underset{\substack{m_i \sim \Poi(p_i\mu) \\ m_j \sim \Poi(p_j\mu)}}{\mathbb{E}}[o_{ii}o_{ij}]
\label{V_2_riscritto}
\end{align}
The expected values which appear in~(\ref{V_2_riscritto})can be easily computed:
\begin{align}
\underset{m_i \sim \Poi(p_i\mu)}{\mathbb{E}} [o_{ii}^2]&=\frac{2(1+2\mu p_i)}{\mu^2}\Tr[\rho_i^2]^2\label{exp_iiii} \\
\underset{\substack{m_i \sim \Poi(p_i\mu) \\ m_j \sim \Poi(p_j\mu)}}{\mathbb{E}}[o^2_{ij}]&=\frac{(\mu p_ip_j(p_i+p_j)+p_ip_j)}{\mu^2}\Tr[\rho_i\rho_j]^2, \qquad i\neq j\label{exp_ijij} \\
\underset{\substack{m_i \sim \Poi(p_i\mu) \\ m_j \sim \Poi(p_j\mu) \\ m_k \sim \Poi(p_k\mu)}}{\mathbb{E}} [o_{ij}o_{ik}]&=\frac{p_ip_jp_k}{\mu}\Tr[\rho_i\rho_j]\Tr[\rho_i\rho_k], \qquad i\neq j\neq k\neq i\label{exp_ijik} \\
\underset{\substack{m_i \sim \Poi(p_i\mu) \\ m_j \sim \Poi(p_j\mu)}}{\mathbb{E}} [o_{ii}o_{ij}]&=\frac{2p_ip_j}{\mu}\Tr[\rho_i\rho_j]\Tr[\rho_i^2] \; \qquad i\neq j\label{exp_iiij}.
\end{align}

Replacing~(\ref{exp_iiii}),~(\ref{exp_ijij}),~(\ref{exp_ijik}) and~(\ref{exp_iiij}) into~(\ref{V_2_riscritto}), { and then isolating the leading order,}  we can conclude that

\begin{align}
V_2&=\sum_{i}\frac{8(1+2\mu p_i)}{\mu^2}(1-p_i)^2\Tr[\rho_i^2]^2+\sum_{i\neq j}\frac{8(\mu p_ip_j(p_i+p_j)+p_ip_j)}{\mu^2}\Tr[\rho_i\rho_j]^2\nonumber\\
&+\sum_{i\neq j\neq k\neq i}\frac{{16}(p_ip_jp_k)}{\mu}\Tr[\rho_i\rho_j]\Tr[\rho_i\rho_k]-\sum_{i\neq j}\frac{32p_ip_j}{\mu}(1-p_i)\Tr[\rho_i\rho_j]\Tr[\rho_i^2]\nonumber\\
&\leq \sum_{i}\frac{16}{\mu}(1-p_i)^2p_i\Tr[\rho_i^2]^2+\sum_{i\neq j}\frac{8 p_ip_j(p_i+p_j)}{\mu}\Tr[\rho_i\rho_j]^2\nonumber\\
&+\sum_{i\neq j\neq k\neq i}\frac{{16}(p_ip_jp_k)}{\mu}\Tr[\rho_i\rho_j]\Tr[\rho_i\rho_k]-\sum_{i\neq j}\frac{32p_ip_j}{\mu}(1-p_i)\Tr[\rho_i\rho_j]\Tr[\rho_i^2]+O(N/\mu^2) \; .\nonumber\\ 
\label{risultato_V2}
\end{align}

\subsection{Bound on $V_1$+ $V_2$}

We start by observing that
\begin{eqnarray}
&0\leq \Tr[(\rho_i\sqrt{\rho_j} - \rho_k\sqrt{\rho_j})^\dagger(\rho_i\sqrt{\rho_j} - \rho_k\sqrt{\rho_j})] \nonumber\\& \implies {\Tr[\rho_i\rho_j\rho_k]+\Tr[\rho_i\rho_k\rho_j]} \leq \Tr[\rho^2_i \rho_j] + \Tr[\rho^2_k \rho_j] \; .\label{AMGM}
\end{eqnarray}

Applying~(\ref{AMGM}) to the sum and summing

\begin{equation}
{\sum_{i\neq j\neq k\neq i}\frac{p_ip_jp_k}{\mu}\Tr[\rho_i\rho_j\rho_k]\leq \sum_{i\neq j}\frac{p_i p_j(1-p_i-p_j)\Tr[\rho_i\rho_j^2]}{\mu}}
\label{AMGM_tracce}
\end{equation}

Combining~(\ref{risultato_V1}),~(\ref{risultato_V2}) and using~(\ref{AMGM_tracce}) we have
\begin{align}
&V_1+V_2= O\left( \frac{N}{\mu^2} \right) + 16\sum_{i}\frac{p_i(1-p_i)^2}{\mu}\Tr[\rho^3_i] +8\sum_{i\neq j} \frac{p_i p_j^2 \Tr[\rho_i\rho_j^2]+p_j p_i^2 \Tr[\rho_j\rho_i^2]}{\mu}\nonumber\\
&-32\sum_{i\neq j} \frac{(1-p_i) p_ip_j}{\mu}\left( \Tr[\rho^2_i\rho_j]\right)+{16}\sum_{i\neq j\neq k\neq i}\frac{p_ip_jp_k}{\mu}(\Tr[\rho_i\rho_j\rho_k])+O(N/\mu^2)\nonumber\\
&\leq O\left( \frac{N}{\mu^2} \right) + 16\left(\sum_{i}\frac{p_i(1-p_i)^2}{\mu}\Tr[\rho^3_i]+\sum_{i\neq j} \frac{p_ip_j[(p_j+1-p_i-p_j) \Tr[\rho_i\rho_j^2]-2(1-p_i)\Tr[\rho_i^2\rho_j]]}{\mu}\right)\nonumber\\
&=O\left( \frac{N}{\mu^2} \right) + \frac{16}{\mu} \sum_{i\neq j}p_ip_j\Tr[((1-p_i)\rho_i)(\rho_i-\rho_j)^2]
\leq \sum_{i\neq j}p_ip_j\Tr[\| (1-p_i)\rho_i \|_\infty (\rho_i-\rho_j)^2]
\nonumber \\
&\leq O\left( \frac{N}{\mu^2} \right) + \frac{16}{\mu}  \sum_{i\neq j}p_ip_j\Tr[(\rho_i-\rho_j)^2]
\nonumber \\ &= O\left( \frac{N}{\mu^2} \right) + \frac{16}{\mu}  \sum_{i\neq j}p_ip_jD^2_{HS}(\rho_i, \rho_j) = O\left( \frac{N}{\mu^2} \right) + \frac{16 \mathcal{M}_{HS}^2}{\mu}.
\end{align}
\end{appendix}
\end{document}